\documentclass[sn-apa, pdflatex]{sn-jnl}

\usepackage{graphicx}%
\usepackage{multirow}%
\usepackage{amsmath,amssymb,amsfonts}%
\usepackage{amsthm}%
\usepackage{mathrsfs}%
\usepackage[title]{appendix}%
\usepackage{xcolor}%
\usepackage{textcomp}%
\usepackage{manyfoot}%
\usepackage{booktabs}%
\usepackage{algorithm}%
\usepackage{algorithmicx}%
\usepackage{algpseudocode}%
\usepackage{listings}%
\usepackage{subcaption}%

\usepackage{caption}%

\newtheorem*{assumption}{Assumption}

\makeatletter%
\def\@acknow{}%
\long\def\EarlyAcknow#1 \par{%
\def\@acknow{\abstractfont\abstracthead*{}%
#1\par}}%

\def\printabstract{\ifx\@acknow\empty\else\@acknow\fi\par%
    \ifx\@abstract\empty\else\@abstract\fi\par}
\makeatother

\begin{document}

\title[Estimating Input Coefficients for Regional Input--Output Tables Using Deep Learning with Mixup]{Estimating Input Coefficients for Regional Input--Output Tables Using Deep Learning with Mixup}

\author*{\fnm{Shogo} \sur{Fukui}}\email{sfukui@yamaguchi-u.ac.jp}

\affil*{\orgdiv{Department of Economics}, \orgname{Yamaguchi University}, \country{Japan}}

\EarlyAcknow{This preprint has not undergone peer review (when applicable) or any post-submission improvements or corrections. The Version of Record of this article is published in Computational Economics, and is available online at \url{https://doi.org/10.1007/s10614-024-10641-1}.}


\abstract{
    An input--output table is important data for analyzing the economic situation of a region.
    Generally, the input--output table for each region (regional input--output table) in Japan is not always publicly available, so it is necessary to estimate the table.
    In particular, various methods have been developed for estimating input coefficients, which are an important part of the input--output table.
    Currently, non-survey methods are often used to estimate input coefficients because they require less data and computation, but these methods have some problems, such as discarding information and requiring additional data for estimation.

    The purpose of this paper is to present a method for estimating input coefficients using artificial neural networks (ANNs) with better accuracy than the conventional non-survey methods.
    To avoid over-fitting due to the small data used, data augmentation, called mixup, was introduced to increase the data size by generating virtual regions through region composition and scaling.

    By comparing the estimates with published values of input coefficients for Japan as a whole, we found that our method was more accurate and stable than some representative non-survey methods.
    The estimated input coefficients for three Japanese cities were generally close to the published values for each city.
}


\maketitle

\section{Introduction}
Input--output tables record the flow of products and services by industry for a given region and time period, and play an important role in various quantitative analyses, such as economic spillover effects analysis and general equilibrium analysis.
In general, input--output tables do not exist for all regions and time periods, since compiling such tables requires a large amount of primary data and a great deal of work.
In the case of Japan, the only input--output table that is strictly derived from primary data is the one for Japan as a whole.
While estimated input--output tables are available for prefectures and some cities, many smaller administrative units (cities, towns, and villages) do not publish their own input--output tables.
Therefore, if an analysis using input--output tables is to be attempted for a small region, it is necessary to estimate input--output tables.

Estimation methods for the input--output table can be classified as survey, non-survey, and hybrid methods.
For small regions, non-survey and hybrid methods are commonly used.
\cite{Richardson1985} summarized specific methods commonly used to estimate regional input--output tables.
The survey method derives an input--output table by synthesizing primary data obtained from surveys of firms and consumers involved in the economy of the target region.
The survey method is highly accurate because it uses information from individual firms and consumers, but it requires a large amount of primary data for estimation. As a result, survey methods are often used only for entire countries and very rarely for small areas such as cities.
The non-survey methods use less data and the calculation method is relatively simple, making it possible to estimate an input--output table at low cost.
The hybrid method estimates the table by combining another survey or data with the results of the non-survey estimation.

The estimation of input coefficients is the most important part of the estimation of input--output tables.
The input coefficient is the intermediate input, which represents the transfer of output between industries, divided by the gross output of each industry.
The coefficients are summarized in a matrix, the input coefficient matrix.
If the input coefficient matrix can be estimated, the intermediate inputs are obtained by the product of the input coefficients and the gross output and thus contributes greatly to the estimation of the entire input--output table.
Even if the full input--output table cannot be estimated, the economic spillover effects are calculated by estimating the input coefficient matrix.

Many previous studies have presented non-survey methods for estimating input coefficients.
The location quotient method(LQ) and the RAS method(RAS) are representative non-survey methods for estimating input coefficients for a region.
According to \cite{Isserman1977}, ``location quotient is the ratio of an industry's share of the economic activity of the economy being studied to that industry's share of another economy.''
LQ uses this location quotient to get the input coefficients of the region to be estimated from the coefficients of the reference region.
The estimation of input coefficients by location quotients has derivatives such as CILQ, RLQ, and FLQ\citep{Flegg2013}.
\cite{Bonfiglio2008} compared the estimation accuracy of methods using the location quotient and found that the accuracy of FLQ and the augmented FLQ(AFLQ) was more accurate than the other methods by constructing a multi-regional input--output table using Monte Carlo simulations.
Even in recent years, new methods have been developed, such as FLQ+ by \cite{Flegg2021}.
RAS estimates the input coefficient matrix by adjusting the initial input coefficient matrix using the bi-proportional method with the total intermediate demands, total intermediate inputs, and total gross outputs of each industry.
RAS is said to have originated in Stone's study\citep{Bacharach1970}.
\cite{Hewings1977} estimated the 1965 input coefficient matrix for Kansas using the RAS method, based on the 1963 input coefficient matrix for Washington, and showed that RAS is highly accurate in its estimation, although caution in estimation and interpretation was required.
The estimation of regional input coefficient matrices by RAS has also been improved and extended by many researchers; for example, \cite{Holy2022} have developed a multidimensional RAS that decomposes an input--output table for a country by region.
\footnote{
    \cite{Lahr2004} summarized several extensions of RAS.
}

These non-survey methods require strong assumptions to attempt estimates from small amounts of data.
There has been a long debate about whether these assumptions are realistic and about the estimation accuracy of non-survey methods relying on these assumptions.
\cite{Round1983} provided a critical perspective on the theoretical aspects of non-survey methods.
As for the empirical problems with non-survey methods, for example, \cite{Riddington2006} measured the economic effects of tourism expenditures using non-survey and hybrid methods for a group of small regions in Scotland and shows that estimates using SLQ and CILQ may lead to erroneous conclusions.
\cite{Szabo2015}, on the other hand, showed that the non-survey method has both theoretical and empirical shortcomings, but still argued that the non-survey method is necessary to efficiently address the lack of data in the region.

Looking at the non-survey method from the perspective of estimation accuracy, the problems are the decrease in accuracy because much information necessary for estimation is discarded, and the instability of accuracy due to the data selected for estimation.
Non-survey methods are used to reduce the time and resources required for estimation.
LQ and RAS, which require less data, are often used to estimate regional input coefficient matrices because of the limited data available for small regions.
However, the small amount of data used for estimation means that much information about the local economy is discarded.
Despite the large amount of primary data required to construct an input--output table, non-survey methods attempt to estimate input coefficients with relatively few numbers.
Such discarding of information can be a factor that reduces the accuracy of the estimation.
In addition, both LQ and RAS must use input coefficient matrices other than the time and region to be predicted, and RAS requires total intermediate demands, total intermediate inputs, and total gross outputs for each industry in the region.
The accuracy of these methods depends on the data used for estimation.

In addition to these non-survey methods, there are other methods for estimating input coefficients by regression. There are fewer previous studies on regression projection methods compared to LQ and RAS.
\cite{Gerking1976} proposed a method to calculate input coefficients as estimates of partial regression coefficients through regression analysis with intermediate input as the objective variable and gross output as the explanatory variable.
Gerking's method aims to avoid the effects of measurement error in intermediate input and gross output data on the estimation results.
In small areas, where these data are generally not available, it is difficult to apply Gerking's method directly.
\cite{Papadas2002} constructed an artificial neural network(ANN) as a forecasting model for the input coefficients, obtained forecasts of input coefficients for 1992 based on 1984 data from the United Kingdom, and attempted to compare them with RAS.
In that study, an ANN was set up with input coefficients as the objective variable and the ratio of intermediate demand to gross output in the input source industry and the ratio of intermediate input to gross output in the input target industry as the two explanatory variables, and the model was trained with 49 observations in 7 industries in the input--output table.
However, the form of the model was limited to two explanatory variables and one intermediate layer, and a single model was applied to all 49 input coefficients.
For these limitations, it may not have achieved predictive accuracy beyond RAS.

Although there are various restrictions and drawbacks to predicting input coefficients by regression, it is expected to alleviate some of the problems associated with current mainstream non-survey methods.
When input coefficients are predicted by regression, a variety of variables representing local economic conditions can be included as explanatory variables.
Therefore, more information can be included in the estimation than in LQ or RAS, and higher prediction accuracy can be expected.
In addition, while the LQ and RAS have restrictions on the variables used in the estimation, regression rarely has such restrictions.

In recent years, deep learning has been developed, where multi-layer ANNs are estimated and used for prediction.
The introduction of deep learning into the forecasting of input coefficients by regression is expected to improve the prediction accuracy.
In the fields of economics and finance, the effectiveness of ANN-based forecasting has been demonstrated in a number of cases\citep{Ramyar2019, Law2019, Abbasimehr2020}.
However, when the data available for model estimation is small, over-fitting can reduce the accuracy of forecasts.
As is clear from the data used in this study and \cite{Papadas2002}, the data available for estimating regional input coefficients are generally small.
Therefore, over-fitting is inevitable when deep learning is applied directly to predict regional input coefficients.

To mitigate the effects of over-fitting, data augmentation, which involves manipulating the original data to increase its size, is widely used in machine learning, especially for data such as images, text, and audio.
A number of studies have demonstrated the effectiveness of data augmentation.
\cite{Dao2019} have shown that data augmentation can be approximated as feature averaging and variance regularization by representing the augmentation process as a Markov chain.
\cite{Wu2020} have shown that when the true model is a linear model, data augmentation by rotation or scaling of image data can reduce the parameter estimation error due to the addition of new information, and data augmentation by linear combination of multiple data can produce a regularization effect.

A relatively new data augmentation method is mixup, which has been introduced by \cite{Zhang2018}.
This is a method that augments data by linearly combining variables for two observations.
Most regional macroeconomic data have the property that the values of several regions can be added together to obtain the value that would be obtained if these regions were considered as one, and that the value of a region can be scaled by multiplying the value of a region by a constant.
From these properties, it is conceivable to augment the data by applying mixup to the regional macroeconomic data.
\footnote
{
From this idea, in a paper published in Japanese in 2021, I have attempted to estimate the input coefficients by applying a partial mixup, focusing only on the additivity of regional data, and have shown that it could predict coefficients for two regions in Japan with the same level of accuracy as RAS.
However, the additivity-only mixup could only use very limited data as explanatory variables to ensure vicinity for regions with different economic sizes.
For example, when using prefecture data to project input coefficients for a city, the sum of quantitative variables such as prefecture income and labor force obtained by mixup is not close to the city.
Therefore, only variables that are less related to the size of prefectures, such as prefecture income per capita and the percentage of the population employed in industry, were allowed as explanatory variables.
Building on the work of the previous research, this study introduces a full mixup that also considers scalar products for regional data, allowing more types of information to be included as explanatory variables.
In addition, I attempt to make more accurate predictions by adding a new procedure for establishing regional vicinity when training the model.
}

In this study, we use mixup of \cite{Zhang2018} to augment data by generating virtual regions from data of some Japanese prefectures and cities, and present a deep learning method to predict input coefficients.
After extending mixup to regional data, we consider the application of ANN with input coefficients as objective variables (Section 2), mixup is performed on the data of Japanese prefectures and cities to train the ANNs with the input coefficients as the objective variables, and the ANNs are used to predict the input coefficients for the three cities of Japan(Section 3).
Finally, we discuss the methods presented in this study (Section 4).

\section{Extension of mixup to regional input coefficients prediction}
Data augmentation is a method of processing the original data to generate new data. In general image recognition, machine learning is performed with the numerical information of the image as the explanatory variable and the label of ``what the image is'' as the objective variable.
The model trained by machine learning can lead to over-fitting if the data size used for training is small. Over-fitting causes the trained model to fit the training data too well, resulting in reduced prediction accuracy for unknown data.
One way to deal with over-fitting is data augmentation, which creates new data from existing data.
In the case of image data, a new image is generated by slightly rotating and scaling the image in the original data, but the labels remain the same as in the original image.
Data augmentation creates new image--label pairs in this way, increasing the size of the data.

We confirm the procedure for data augmentation by mixup based on \cite{Zhang2018}.
For the $i$-th individual in the data of size $n$ $(i = 1, \ldots, n)$, $\boldsymbol{x}_i$ is the value of the explanatory variable, $y_i$ is the value of the objective variable, and the original training data are $(\boldsymbol{x}_i, y_i)$.
In mixup, a new individual $(\bar{\boldsymbol{x}}, \bar{y})$ is generated from two individuals $(\boldsymbol{x}_A, y_A)$ and $(\boldsymbol{x}_B, y_B)$, randomly selected from the training data, as follows:
\begin{align}
    \bar{\boldsymbol{x}} &= \lambda \boldsymbol{x}_A + (1-\lambda) \boldsymbol{x}_B \\
    \bar{y} &= \lambda y_A + (1-\lambda) y_B
\end{align}
where $\lambda \in [0,1]$ and $\lambda \sim \text{Beta}(\alpha, \alpha)$.
When mixup is performed, the random number generated from this beta distribution is used as $\lambda$.

Let $\boldsymbol{x}_i$ be the numerical density of each pixel in the grayscale image and $y_i$ the label of that image.
Now suppose that two images $(\boldsymbol{x}_1, y_1)$ and $(\boldsymbol{x}_2, y_2)$ are chosen at random.
Mixup produces a new composite image by diluting image $\boldsymbol{x}_1$ to $\lambda \times 100$\% and image $\boldsymbol{x}_2$ to $(1-\lambda) \times 100$\%.
The resulting image can be considered as the first image with probability $\lambda$ and the second image with probability $(1 - \lambda)$.
According to \cite{Zhang2018}, mixup is equivalent to vicinal risk minimization\citep{Chapelle2000} with a certain generic vicinity distribution for each observation $(\boldsymbol{x}_i, y_i)$.
In addition, \cite{Zhang2021} showed that the loss function in mixup contains regularization terms.

Data augmentation is a method of generating data based on prior information or knowledge.
In conventional data augmentation, which transforms a single image data, the prior knowledge is the invariance of the image data, i.e., the labels are unchanged with respect to image rotation and scaling.
In mixup, the prior knowledge is that ``linear interpolation of feature vectors should lead to linear interpolation of the associated targets.''\citep{Zhang2018}.

\begin{figure}[t]
    \begin{minipage}[b]{0.58\columnwidth}
        \centering
        \includegraphics[width=\columnwidth, clip]{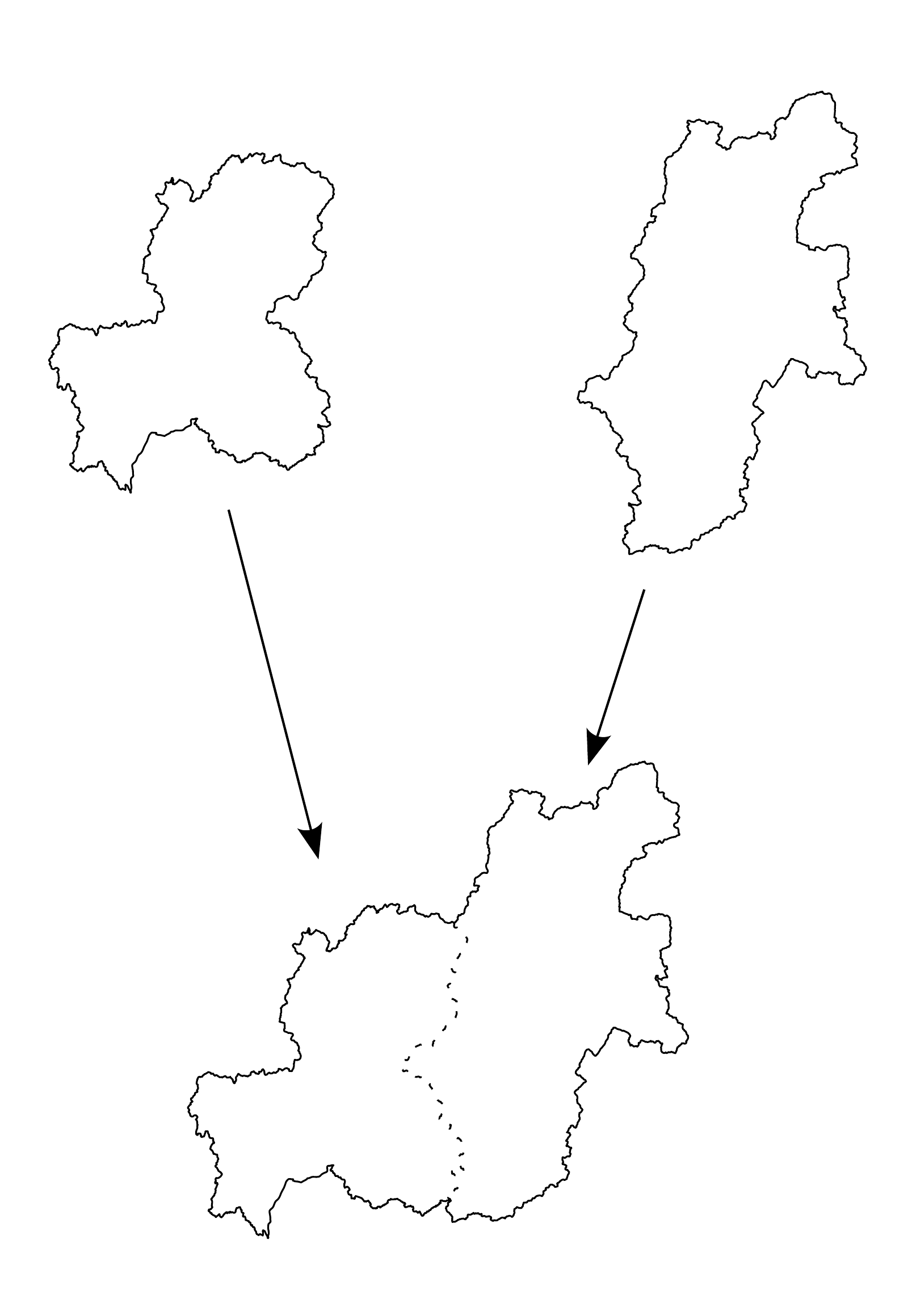}
        \subcaption{Composition}\label{Fig:Composition}
    \end{minipage}
    \hspace{0.04\columnwidth}
    \begin{minipage}[b]{0.3\columnwidth}
        \centering
        \includegraphics[width=\columnwidth, clip]{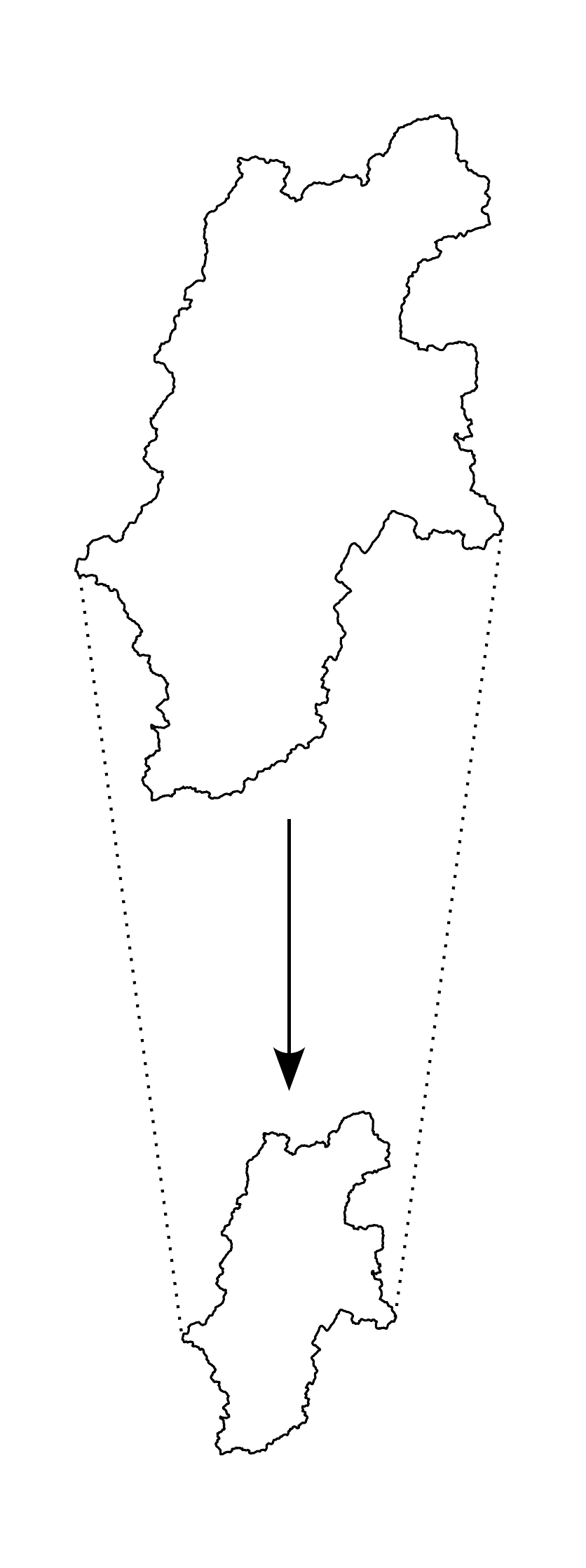}
        \subcaption{Scaling}\label{Fig:Scaling}
    \end{minipage}
    \caption{Images of the composition and scaling of regions. Maps of Nagano and Gifu prefectures in Japan are shown as examples, which are created by processing the ``Digital National Land Information(administrative area data)'' of the Ministry of Land, Infrastructure, Transport and Tourism (\url{https://nlftp.mlit.go.jp/ksj/index.html})}
\end{figure}

Consider applying mixup to quantitative macroeconomic variables for regions.
Mixup cannot be applied to all types of data.
For example, it is difficult to adapt the prior knowledge that mixup assumes for qualitative values such as gender and occupation, and for rating scales such as technical support satisfaction.
However, for many quantitative regional macroeconomic variables, operations such as vector sums and scalar products of variables can be meaningful.
For example, the total population of countries in North America, such as the United States, Canada, and Mexico, is equal to the population of all of North America. The same is true for quantitative variables such as income and number of establishments.
That is, if $\boldsymbol{r}_k$ is a vector consisting of quantitative economic variables for region $k(k=1,\ldots,K)$, then the value of the vector when these regions are considered as one can be expressed as $\sum_k \boldsymbol{r}_k$. In other words, the vector sum of quantitative economic variables for each region corresponds to a hypothetical composite of those regions(Figure \ref{Fig:Composition}).

A scalar product for a quantitative variable implies a scaling of values.
If the population of North America is $r$, half of it can be calculated as $0.5 \times r$.
For a vector $\boldsymbol{r}$ consisting of quantitative economic variables,
$\lambda \boldsymbol{r}$ multiplies all the quantitative variables in $\boldsymbol{r}$ by $\lambda$.
This operation implies a hypothetical expansion or contraction of the region(Figure \ref{Fig:Scaling}).

A new virtual region can be generated by linear interpolation, which combines a set of regions by scaling each of them.
The observation vector of a virtual region is generated by the following equation.
\begin{equation}
    \bar{\boldsymbol{r}} = \sum \lambda_k \boldsymbol{r}_k
\end{equation}
where $\boldsymbol{r}_k \ (k = 1, \ldots, K)$ is the vector of observations and $\lambda_k, \ (k = 1, \ldots, K)$ is the constant that scales them for the region $k$.

From the observation of a virtual region obtained by the linear interpolation described above, the explanatory and objective variables of the model are calculated.
In the case of quantities, the value itself contained in $\bar{\boldsymbol{r}}$ becomes the value of these variables.
When indices or ratios are used as explanatory and objective variables, the quantities that are the source of the indices and ratios are included in $\boldsymbol{r}_k$ and the values of these variables are calculated after linear interpolation.
For example, if we want to use the unemployment rate as an explanatory variable, we can include the number of unemployed and the labor force in $\boldsymbol{r}_k$ and calculate the unemployment rate from data for a virtual region derived by linear interpolation.
In the case of an input--output table of competitive import type, a region's intermediate inputs include inputs from other regions, but as explained in the appendix, the sum of a group of regions' intermediate inputs is equal to the intermediate inputs of the combined group of these regions measured as a single region.
The scalar product of a region's intermediate inputs is equal to the region's intermediate inputs when the region is scaled and measured as a single region.
These properties also hold for the gross output.
\footnote{It is shown in the appendix that the sum of the gross output of a group of regions is equal to the gross output of the regions when they are considered as a single region.}
Therefore, it is possible to calculate the input coefficient of a virtual region obtained by linear interpolation as the ratio of the intermediate input to the gross output.

In order to derive the prior knowledge that is the prerequisite for mixup, we make the following assumption.
\begin{assumption}
Let the input coefficient $a_{i,j}$ from industry $i$ to $j$ be the objective variable and $\boldsymbol{x}^* = (x_1, \ldots, x_m)^\prime$ be the explanatory variable.
Then $a_{i,j} = f_{i,j}^*(\boldsymbol{x}^*)$ holds uniquely for all regions.
\end{assumption}
This is similar to the assumption that the usual econometric model assumes one regression equation for all observed individuals, which means that the input coefficient $a_{i,j}$ is determined by only one function $f_{i,j}^*$ for all regions, including the virtual region generated by linear interpolation.
The values in the input--output table are calculated from the various primary data according to predefined rules.
In principle, this rule is the same for all regions, so it is natural to make the above assumptions for the input coefficient predictions.
With this assumption, if we denote the values of the primary data as $\boldsymbol{x}^*$ and the rule for calculating the values of the input--output table from the primary data as the function $\boldsymbol{F}^*$, then the input--output table can be expressed as $\boldsymbol{F}^*(\boldsymbol{x}^*)$.
Restricting to input coefficients, if $f_{i,j}^*$ is the rule for computing $a_{i,j}$ from primary data, then
\begin{equation}
    a_{i,j} = f_{i,j}^* (\boldsymbol{x}^*). \label{Eq:PriorKnowledge}
\end{equation}

Based on the above assumption, for a hypothetical region $v$ generated by linear interpolation, if its primary data vector is $\boldsymbol{x}^*_v$, the input coefficient can be calculated as $f_{i,j}^*(\boldsymbol{x}^*_v)$. On the other hand, the input coefficient $a^v_{i,j}$ for the region $v$ has already been obtained by mixup.
Therefore, $a^v_{i,j} = f_{i,j}^*(\boldsymbol{x}^*_v)$.
The equation of (\ref{Eq:PriorKnowledge}) is nothing more than the prior knowledge in this analysis.
In other words, mixup can be applied to the estimation of input coefficients by changing the prior knowledge of the original mixup as follows:
For a virtual region obtained by linear interpolation, the feature vectors should lead to the associated target.

The regional composition and scaling in the above mixup do not represent actual regional merger and division.
If two cities are actually merged, the values of the economic variables after the merger should be different from the sum of the values before the merger due to changes in the economic structure.
Alternatively, even if a city is divided into two parts so that the area is divided equally, this does not necessarily mean that the population and income are also divided in half.
Composition and scaling in mixup simply derive ``the value of the data when individual areas are measured together'' and ``the value of the data when an area is scaled by a constant,'' and never assumes any change in economic structure.

In this study, $f_{i,j}^*$ is approximated by a multi-layer ANN.
Generally, in small areas such as cities, the primary data needed to compute input--output tables are not sufficiently measured.
In this case, the original explanatory variable $\boldsymbol{x}^*$ must be replaced by another variable $\boldsymbol{x}$ derived from the available data.
Similarly, the true function $f_{i,j}^*$ is approximated by another function $f_{i,j} (\boldsymbol{x})$.
Since the purpose of this study is to predict input coefficients in a small region with high accuracy, a multi-layer ANN is set up as $f_{i,j} (\boldsymbol{x})$.

\section{Empirical analysis for Japan}
\begin{figure}[t]
    \centering
    \includegraphics[width=0.9\columnwidth, clip]{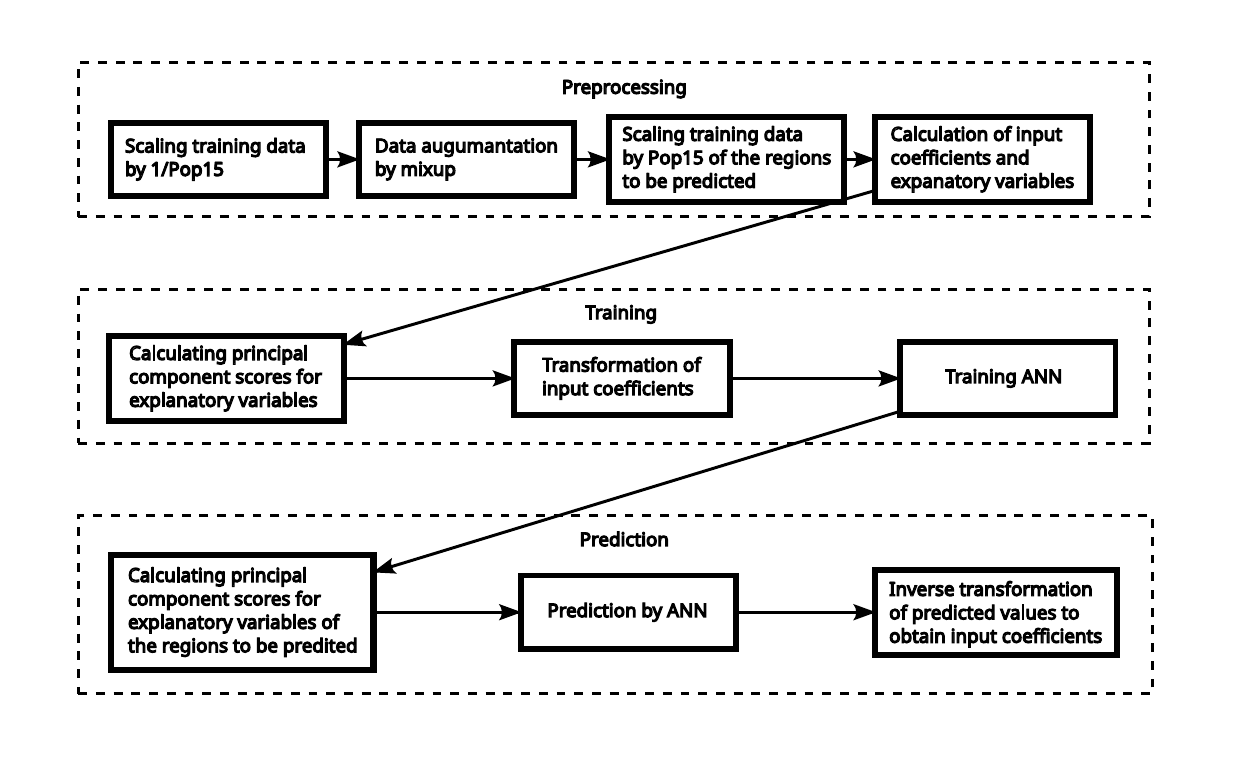}
    \caption{Model training and inference flow} \label{Fig:Chart}
\end{figure}

In the following, after data augmentation by mixup for some prefectures and ordinance-designated cities in Japan, deep learning is performed using the input coefficients as the objective variable.
We will check the prediction accuracy of the trained model for the input coefficients for Japan as a whole and predict the input coefficients for some cities in Japan.
Figure \ref{Fig:Chart} shows the flow of training and inference, including data pre- and post-processing.
\footnote{F\# is used for Mixup and other data processing, and Tensorflow is used for deep learning and prediction in Python.}

The data used in this analysis is presented in Table \ref{Tab:OrigData}.
\footnote{
    All data used in this study are publicly available from public institutions and can be obtained from the sources listed in Table \ref{Tab:OrigData}.
    Links to the input--output table for each region are compiled on the the Pacific Rim Association for Input-Output Analysis(PAPAIOS) website(\url{http://www.gakkai.ne.jp/papaios/en/io_j.html}).
}
All input--output tables used in the analysis are of the competitive import type.
Data from 2015 or the most recent year of 2015 are used to predict input coefficients for 2015.
Data containing non-numeric values are excluded before training the model.
The regions for which all of these data are currently available are included in the analysis(Table \ref{Tab:Areas}).
Regions containing outliers are excluded from the analysis, even if all data are available.
In Japan's administrative divisions, the entire country is divided into 47 prefectures, and each prefecture is further divided into a number of cities, towns, and villages.
The names of cities in Table \ref{Tab:Areas} are appended with the prefecture in which they are located.
Among the regions in Table \ref{Tab:Areas}, the group of regions that are the target of the forecast (Japan as a whole, Gujo City in Gifu Prefecture, Sapporo City in Hokkaido, and Okayama City in Okayama Prefecture) are excluded to obtain the values of the input coefficients and explanatory variables in Table \ref{Tab:Variables} by mixup.
The input coefficients are calculated after recompiling the input--output tables for each region so that there are 12 industry sectors as shown in Table \ref{Tab:Industries}.
For an input coefficient $a_{i, j}$, $i$ and $j$ indicate the value of ``Order'' from Table \ref{Tab:Industries}.
For example, $a_{1,2}$ represents the input coefficient from agriculture to mining.

\begin{table}[tb]
    \centering
    \caption{The data set used in this study. The minor classification refers to the 590 industry classifications of the Economic Census, and the large classification refers to the 19 industry classifications of the Economic Census}\label{Tab:OrigData}
    \begin{tabular}{lll}
        \toprule
        Variable name & Data & Source \\
        \midrule
        $\text{SFirm}_k$ & Number of establishments & 2014 Economic Census \\
         & (minor classification) & for Business Frame \\
        $\text{SEmp}_k$ & Number of employees & Ibid. \\
         & (minor classification) & \\
        $\text{VA}_k$ & Added value(large classification) & 2012 Economic Census \\
         & & for Business Activity \\
        $\text{Sales}_k$ & Sales(large classification) & Ibid. \\
        $\text{Firm}_k$ & Number of establishments & Ibid. \\
         & (large classification) & \\
        $\text{Income}_k$ & Taxable income & Statistical Observations \\
         & & of Prefectures 2015 and \\
         & & Statistical Observations \\
         & & of Municipalities 2015 \\
        $\text{TP}_k$ & Taxpayer & Ibid. \\
        $\text{PopLF}_k$ & Population in labor force & Ibid. \\
        $\text{Unemp}_k$ & Number of unemployed persons & Ibid. \\
        $\text{Pop15}_k$ & Total population(15 and over) & Calculated by the author from \\
        & & Statistical Observations of \\
        & & Prefectures 2015 and Statistical \\
        & & Observations of municipalities 2015 \\
        $A_{i,j,k}$ & Intermediate input(12 industries) & Calculated by the author from \\
         & & the input--output table for each \\
         & & prefecture and city. \\
        $Y_{j,k}$ & Gross output(12 industries) & Ibid. \\
        \bottomrule
    \end{tabular}
\end{table}

\begin{table}[tb]
    \centering
    \caption{Target areas for analysis. City names are followed by the name of the prefecture to which the city belongs} \label{Tab:Areas}
    \begin{tabular}{ll}
        \toprule
        \multicolumn{2}{l}{For training} \\
        Prefectures & Hokkaido, Aomori, Iwate, Miyagi, Akita, Yamagata, Fukushima, Ibaraki, \\
          & Tochigi, Gunma, Saitama, Chiba, Tokyo, Kanagawa, Niigata, Toyama,  \\
          & Yamanashi, Nagano, Gifu, Shizuoka, Aichi, Mie, Shiga, Kyoto, Osaka, Hyogo,  \\
          & Wakayama, Shimane, Okayama, Hiroshima, Yamaguchi, Tokushima, Kagawa,  \\
          & Ehime, Kochi, Fukuoka, Saga, Nagasaki, Kumamoto, Oita, Miyazaki, Kagoshima \\
        Cities & Saitama(Saitama), Yokohama(Kanagawa), Kawasaki(Kanagawa),\\
        &  Fukuoka(Fukuoka)\\
        \midrule
        \multicolumn{2}{l}{For inference} \\
         & Japan, Gujo(Gifu), Sapporo(Hokkaido), Okayama(Okayama) \\
        \bottomrule
    \end{tabular}
\end{table}

\begin{table}[tb]
    \centering
    \caption{Variables used for analysis}\label{Tab:Variables}
    \begin{tabular}{ll}
        \toprule
        Name & Definition \\
        \midrule
        Input coefficient & $=A_{i,j,k} / Y_{j,k}$ \\
        Number of establishments(minor classification) & $=\text{SFirm}_k$ \\
        Composition ratio of the number of & $=\text{SFirm}_k / \sum_k \text{SFirm}_k$ \\
        establishments(minor classification) &  \\
        Number of employees(minor classification) & $=\text{SEmp}_k$ \\
        Composition ratio of the number of  & $=\text{SEmp}_k / \sum_k \text{SEmp}_k$ \\
        employees(minor classification) & \\
        Added value(large classification) & $=\text{VA}_k$ \\
        Added value per firm(large classification) & $=\text{VA}_k / \text{Firm}_k$ \\
        Sales(large classification) & $=\text{Sales}_k$ \\
        Sales per firm(large classification) & $=\text{Sales}_k / \text{Firm}_k$ \\
        Taxable income & $=\text{Income}_k$ \\
        Taxable income per taxpayer & $=\text{Income}_k / \text{TP}$ \\
        Population in labor force & $=\text{PopLF}_k$ \\
        Labor force population ratio & $=\text{PopLF}_k / \text{Pop15}_k$ \\
        Unemployment rate & $=\text{Unemp}_k / \text{PopLF}_k$ \\
        \bottomrule
    \end{tabular}
\end{table}

\begin{table}[tb]
    \centering
    \caption{Industry classification}\label{Tab:Industries}
    \begin{tabular}{ll}
        \toprule
        Order & Industry \\
        \midrule
        1 & Agriculture(agriculture, forestry, and fisheries) \\
        2 & Mining \\
        3 & Manufacturing \\
        4 & Construction \\
        5 & Energy(electricity, gas, and water) \\
        6 & Trade \\
        7 & Finance(finance, insurance, and real estate) \\
        8 & Transportation(transportation and postal) \\
        9 & Communication(information and communication) \\
        10 & Public business \\
        11 & Services \\
        12 & Other industry \\
        \bottomrule
    \end{tabular}
\end{table}

If mixup is performed directly on the above data, the generated data is likely to be concentrated in the vicinity of the prefectures.
As a result, the generated virtual regions will be similar to the prefectures and dissimilar to Japan as a whole and the cities.
Since prefectures tend to have larger individual data values than cities, the generated regional data will be closer to the prefecture values than to the city values if mixup is performed for a prefecture and a city.
As shown in Table \ref{Tab:Areas}, there are more target regions in prefectures than in cities, and the learning results from the data generated by direct mixup will strongly reflect the situation in prefectures.

In the present analysis, the sizes of all regions are standardized based on a single variable before mixup, and the generated regions are converted to the size of the region to be predicted.
Before performing mixup, a scalar product with ($1/\text{Pop15}$) is obtained for the observed values for each prefecture and city, so that the population aged 15 and over of the areas is 1.
After performing mixup on these data, the generated virtual regions are expanded so that the value of the population aged 15 and over is close to the level of the regions to be forecasted.
For the projection of the input coefficients for Japan as a whole, a scalar product is obtained for each observation obtained by mixup, with the $\text{Pop15}$ of Japan as a constant.
To predict the input coefficients for a city, we first set up a uniform distribution with the minimum and maximum of $\text{Pop15}$ in all Japanese cities as the lower and upper bounds.
Then, each time a data set is created in mixup, a scalar product is obtained for the data set using a random number generated from the uniform distribution as a constant.
These converted data sets are used for training.

In the current mixup, two to five regions are randomly selected to generate data for a virtual region.
Unlike the original mixup, the current mixup is not limited to two observations to combine.
In the original mixup, $\lambda$ was assumed to follow a beta distribution $B(\alpha, \alpha)$, whereas in this method, $(\lambda_1, \ldots, \lambda_K)$ is assumed to follow the Dirichlet distribution.
It is also assumed that the $K$ parameters of the Dirichlet distribution have the same value $\alpha$.
The number of regions($=K$) for mixup is the random number generated from a discrete uniform distribution with a lower bound of 2 and an upper bound of 5.
Generating data from many regions is expected to improve the accuracy of extrapolation because the data will be different from the original set of regions.
However, when mixup for a very large number of regions was performed in the preliminary analysis, the accuracy of extrapolation by the trained model tended to decrease.
This may be due to the fact that when too many regions are composited, the features of the generated data are homogenized.
Therefore, to prevent the number of target regions from becoming too large, the maximum value is set to 5, and the number of regions is chosen randomly.
In the process of mixup, prefectures and cities that are in an inclusion relationship are not selected at the same time.
For example, since the city of Sapporo is included in Hokkaido, we do not select a group of regions that includes these two regions.
This is because it is not possible to get the sum of these areas for values that include transfers, such as intermediate input and gross output.

The values of the input coefficients and their explanatory variables are calculated from the data set generated by mixup.
The calculation methods are summarized in Table \ref{Tab:Variables}.
As shown in Figure \ref{Fig:Chart}, when training the model, the principal component scores of the explanatory variables are calculated and these are used as inputs to the model.
From the cumulative contribution ratios, the 50 principal component scores with the highest ratio are used as input variables.
Some of the input coefficients are so small that the derivative calculated by backpropagation may be close to 0.
Therefore, the following transformation, similar to standardization, is performed on the input coefficients during deep learning:
\begin{align}
    \hat{a}_{i,j} &= \frac{a_{i,j} - a_{\text{L}}}{a_{\text{U}} - a_{\text{L}}} \label{Eq:ObjVariableConversion} \\
    a_{\text{L}} &= \max(0, a^{\text{min}}_{i,j}) \notag \\
    a_{\text{U}} &= \min(1, a^{\text{max}}_{i,j} + 0.5(a^{\text{max}}_{i,j} - a^{\text{min}}_{i,j})) \notag
\end{align}
where $a^{\text{min}}_{i,j}$ and $a^{\text{max}}_{i,j}$ are the minimum and maximum values of $a_{i,j}$, respectively, in the training data.
This transformation uses $a^{\text{max}}_{i,j} + 0.5(a^{\text{max}}_{i,j} - a^{\text{min}}_{i,j})$ as a candidate for the maximum value.
This is because it tends to be slightly more accurate than the transformation using $a^{\text{max}}_{i,j}$ as the candidate in predicting input coefficients for Japan as a whole, as described below.
The trained model predicts $\hat{a}_{i,j}$. The $\hat{a}_{i,j}$ is transformed back to get the estimated value of $a_{i,j}$.

\begin{figure}[tb]
    \centering
    \includegraphics[width=0.9\columnwidth, clip]{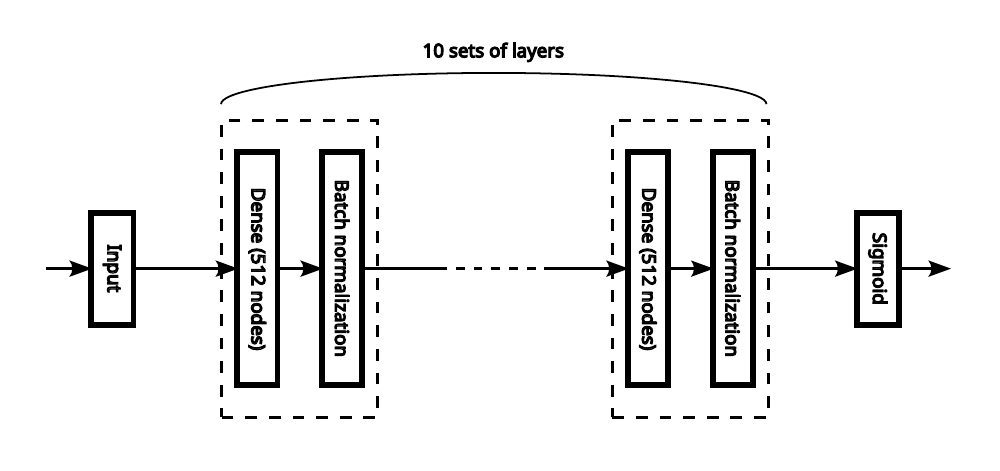}
    \caption{The multi-layer ANN in this analysis} \label{Fig:ANN}
\end{figure}

The multi-layer ANN used for this training is shown in Figure \ref{Fig:ANN}.
It has a relatively standard shape for an ANN: a fully connected layer consisting of 512 nodes and a batch normalization layer as a single pair, and 10 pairs of these connected as an intermediate layer.
In the output layer, $\hat{a}_{i,j}$ is obtained by feeding the linear combination of the intermediate layer outputs into a sigmoid function.
In the fully connected layers of the intermediate layer, the activation function is set as exponential linear unit(ELU), and the initial values of the parameters are determined by the method in \cite{He2015}.
L2 regularization with a hyperparameter of 0.01 is applied to all weight parameters of the fully-connected layers.
The learning rate is varied exponentially and cyclically with a minimum and initial value of 0.0001, a maximum value of 0.01, and a step size of 50\citep{Smith2017}.
When training the model, the parameters are learned by stochastic gradient descent with the least squares error as the loss function and the mini-batch size set to 32.
To speed up the optimization, the parameters are updated using Nesterov Accelerated Gradient(NAG) with a momentum of 0.9.
\footnote{
    NAG has its origins in a 1983 paper in Russian by Yurii E. Nesterov. For more information on NAG, see, for example, \cite{Botev2017}.
}
The step size for optimization is set to a variable value (the maximum number is 200) with early stopping.
In other words, the optimization process is terminated when the least-squares error from the validation data increases for 10 consecutive steps, and the model before the least-squares error starts to increase is used as the training result.
\footnote{These settings in this model training are based on \cite{Geron2019}.}

The specific analysis procedure is described below.
After transforming the original data using a scalar product with ($1/\text{Pop15}$) as a constant, mixup is used to generate data for 50000 regions.
The $\alpha$ of the Dirichlet distribution is set to 1 according to the prediction accuracy in the preliminary analysis.
This makes the Dirichlet distribution a multivariate uniform distribution.
The generated data are multiplied by the value of the population aged 15 or older in the target regions to be predicted to obtain data for virtual regions that are vicinal to the targets.
From these data, the input coefficient, which is the objective variable of the model, and the values of each explanatory variable are calculated.
For the input coefficient, a transformation of (\ref{Eq:ObjVariableConversion}) is performed, and for the explanatory variables, principal component scores are computed to be used as inputs to the model.
Of the data generated by mixup, 40,000 are used for training, 10,000 for testing, and another 20 percent of the training data is used for validation.
Using the training data, the ANN shown in Figure \ref{Fig:ANN} is trained to obtain a prediction model for the input coefficients.
For each region to be predicted, $\hat{a}_{i,j}$ is calculated from the prediction model using the principal component scores of the explanatory variables as input, and the predicted value of the input coefficient ($a_{i,j}$) is calculated by performing the inverse transformation of (\ref{Eq:ObjVariableConversion}).

The above model training and inference are performed for each input coefficient.
In this study, industries are classified into 12 categories, so the number of input coefficients to be covered is $12 \times 12 = 144$.
However, input coefficients that are all 0 in all prefectures and cities in the original data are not included in the training, and the final predicted value is considered to be 0.
Therefore, 131 input coefficients are actually predicted.

First, we check the accuracy of the ANN-based predictions for the input coefficients for Japan as a whole.
The input--output table for Japan as a whole is estimated with relatively high accuracy using the survey method. By comparing with these estimates of the input coefficients, we confirm the accuracy of the method of this study.
Table \ref{Tab:ErrorsJP} shows the prediction accuracy indices derived from the difference between the predicted and actual values of the input coefficients, with smaller values indicating higher accuracy.
\begin{table}[tb]
    \centering
    \caption{Prediction errors of input coefficients for Japan as a whole}\label{Tab:ErrorsJP}
    \begin{tabular}{lcccccccc}
        \toprule
        & ANN & \multicolumn{3}{c}{FLQ} & \multicolumn{3}{c}{RAS(based on} & RAS(based \\
        &  &  &  &  & \multicolumn{3}{c}{prefectures)} & on 2011) \\
        &  & Min & Mean & Max & Min & Mean & Max & \\
        \midrule
        STPE & 0.0434 & 0.3527 & 1.0329 & 3.1693 & 0.0679 & 0.1697 & 0.2829 & 0.1105 \\
        MAD & 0.0017 & 0.0136 & 0.0398 & 0.1221 & 0.0026 & 0.0065 & 0.0109 & 0.0043 \\
        U2 & 0.0468 & 0.4615 & 1.7153 & 5.4916 & 0.0750 & 0.2257 & 0.3913 & 0.1148 \\
        RMSE & 0.0035 & 0.0344 & 0.1279 & 0.4094 & 0.0056 & 0.0168 & 0.0292 & 0.0086 \\
        MAPE & 0.0764 & 0.3519 & 0.9990 & 3.7559 & 0.1162 & 0.3270 & 0.8963 & 0.1940 \\
        \bottomrule
    \end{tabular}
\end{table}
The indices in Table \ref{Tab:ErrorsJP} refer to \cite{Hosoe2013} and are calculated using the following equations:
\begin{align*}
    \text{STPE} &= \sum_{i,j} \left| \tilde{a}_{i,j} - a_{i,j} \right| / \sum_{i,j} a_{i,j} \\
    \text{MAD} &= \sum_{i,j} \left| \tilde{a}_{i,j} - a_{i,j} \right| / N_a \\
    \text{U}_2 &= \sqrt{ \sum_{i,j} \left( \tilde{a}_{i,j} - a_{i,j} \right)^2} / \sqrt{\sum_{i,j} a^2_{i,j}} \\
    \text{RMSE} &= \sqrt{\left[ \sum_{i,j} \left( \tilde{a}_{i,j} - a_{i,j} \right)^2 \right] / N_a} \\
    \text{MAPE} &= (1 / N_a) \sum_{i,j} \left| (\tilde{a}_{i,j} - a_{i,j}) / a_{i,j} \right|
\end{align*}
where $a_{i,j}$ is the published value of the input coefficient and $\tilde{a}_{i,j}$ is its estimated value.
The ``ANN'' column is the prediction accuracy in deep learning.
Each column of ``FLQ'' shows the prediction accuracy of $\tilde{a}_{i,j}$ in the following estimation equation from \cite{Flegg2021}:
\begin{align*}
    a^r_{i,j} &=
    \begin{cases}
        \tilde{a}_{i,j} \text{FLQ}_{i,j} &\quad \text{FLQ} < 1 \\
        \tilde{a}_{i,j} &\quad \text{FLQ} \ge 1
    \end{cases} \\
    \text{FLQ}_{i,j} &=
    \begin{cases}
        \lambda (x^r_i / x^n_i) / (x^r_j / x^n_j) &\quad i \neq j \\
        \lambda (x^r_i / x^n_i) / (x^r / x^n) &\quad i = j
    \end{cases} \\
    \lambda &= \left[ \log_2 (1 + (x^r / x^n)) \right]^2.
\end{align*}
In the inference by FLQ, $x^r_i$ is the gross output of industry $i$ in region $r$, $x^n_i$ is the gross output of industry $i$ nationwide, $x^r$ is the gross output in region $r$, and $x^n$ is the gross output nationwide. We also set $\delta = 0.1$.
From this equation, the predicted value of the input coefficients in the national input--output table is calculated as $\tilde{a}_{i,j}$ when the actual input coefficients for each of the 42 prefectures are given for $a^r_{i,j}$, and the prediction accuracies are displayed in Table \ref{Tab:ErrorsJP}.
Each column of ``RAS'' is the accuracy of the input coefficients estimated by RAS.
The results of RAS are shown for the case based on the input coefficients for each prefecture in 2015, and for the case based on the input coefficients for Japan as a whole in 2011.
Note that the minimum, average, and maximum accuracies for FLQ and RAS are listed side by side because the prediction results differ depending on the prefectural data used as reference.
In addition, since RAS uses actual values from the 2015 national input--output table for intermediate inputs, intermediate demand, and gross output, their forecast errors, which may occur in practice, are zero and do not affect these prediction accuracies of the input coefficients.

From Table \ref{Tab:ErrorsJP}, the accuracy of prediction of input coefficients by deep learning for Japan as a whole is higher and more stable than that of FLQ and RAS.
In Table \ref{Tab:ErrorsJP}, the prediction errors of deep learning are smaller than those of FLQ and RAS.
In other words, the method presented in this study can predict the national input coefficients well.
In addition, FLQ and RAS have different prediction errors depending on the reference input coefficients, but deep learning does not cause such fluctuations in these errors.
Moreover, the errors of deep learning are still lower than those of RAS, which uses actual values for intermediate inputs, intermediate demands, and gross outputs.

We then examine the prediction accuracy for city-level input coefficients.
Unlike national input--output tables, city-level input--output tables are typically inferred by hybrid or non-survey methods, which have larger inference errors for true input--output tables than survey methods.
Therefore, it is difficult to rigorously measure the accuracy of input coefficient forecasting methods for cities.
In the following, I obtain predictions of input coefficients using deep learning for three cities (Gujo, Sapporo, and Okayama) and discuss their characteristics from the errors against the published input coefficients.
Gujo City has a smaller economic scale compared to the ordinance-designated cities such as Sapporo and Okayama.
To confirm the nature of the predicted values for cities with smaller scales, the accuracy of the extrapolation is checked for Gujo City.
For Sapporo and Okayama City, they are randomly selected from among the ordinance-designated cities as targets for extrapolation.

Table \ref{Tab:ErrorsCity} displays the errors relative to the published values for the input coefficients for each city as in Table \ref{Tab:ErrorsJP}.
\begin{table}[tb]
    \centering
    \caption{Prediction errors of input coefficients in the three cities}\label{Tab:ErrorsCity}
    \begin{tabular}{lllllll}
        \toprule
        & \multicolumn{2}{c}{Gujo} & \multicolumn{2}{c}{Sapporo} & \multicolumn{2}{c}{Okayama} \\
        & ANN & RAS & ANN & RAS & ANN & RAS \\
        \midrule
        STPE & 0.2508 & 0.2753 & 0.2971 & 0.2728 & 0.2422 & 0.1811 \\
        MAD & 0.0104 & 0.0104 & 0.0116 & 0.0097 & 0.0097 & 0.0066 \\
        $U_2$ & 0.3156 & 0.5190 & 0.4190 & 0.3820 & 0.3335 & 0.1991 \\
        RMSE & 0.0234 & 0.0368 & 0.0298 & 0.0259 & 0.0239 & 0.0136 \\
        MAPE & 0.3959 & 0.0818 & 0.8203 & 0.6214 & 0.5141 & 0.2884 \\
        \bottomrule
    \end{tabular}
\end{table}
The forecast error of RAS is shown as a comparison with the forecast of ANN.
In the inference of RAS, the input coefficients of the prefecture in which each city is located are used as the initial values, and the actual estimates published by each city are used for the total intermediate demands, total intermediate inputs, and total gross outputs.

From Table \ref{Tab:ErrorsCity}, the prediction accuracy varies depending on the city.
While the error values of ANN are smaller in Gujo except for MAPE, the error values of RAS are generally smaller in Sapporo and Okayama.
However, the published estimates of total intermediate demands, total intermediate inputs, and total gross outputs are used in the calculation of RAS.
It should be noted that the actual forecasts of RAS include these forecast errors.

The prediction error of these cities by the deep learning is checked for each input coefficient.
\begin{figure}[tb]
    \centering
    \begin{minipage}[c]{0.95\hsize}
        \centering
        \includegraphics[width=0.9\columnwidth, clip]{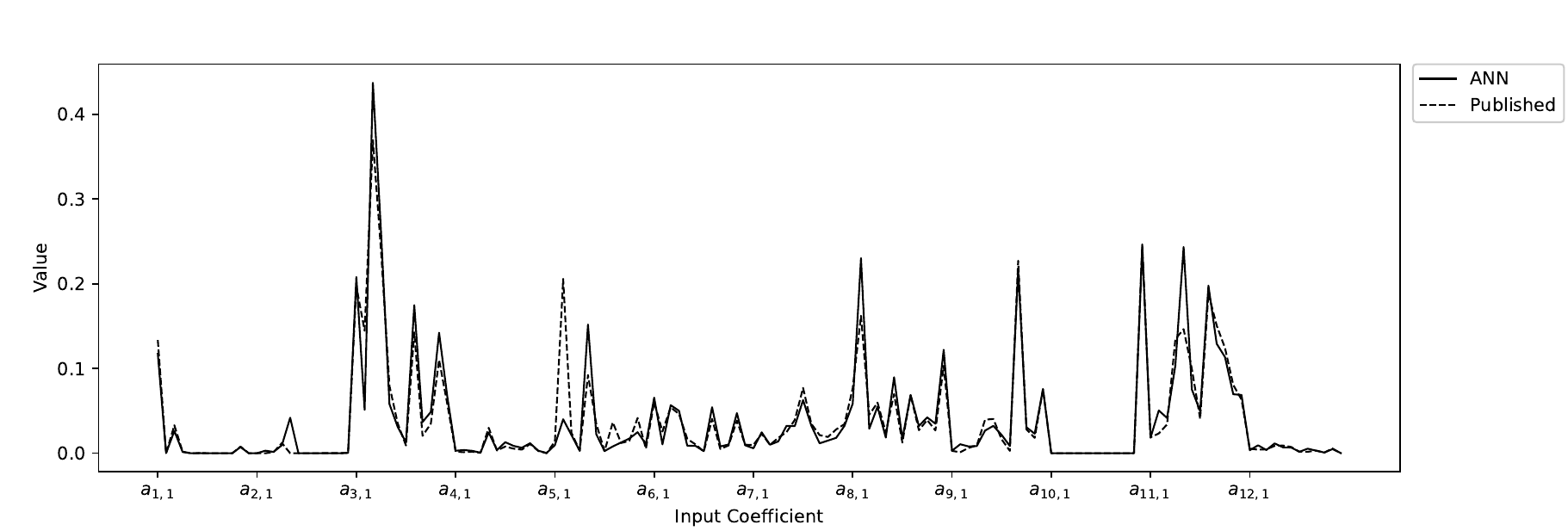}
        \subcaption{Input coefficients of Gujo}\label{Fig:Forecasts21219}
    \end{minipage}\\
    \begin{minipage}[c]{0.95\hsize}
        \centering
        \includegraphics[width=0.9\columnwidth, clip]{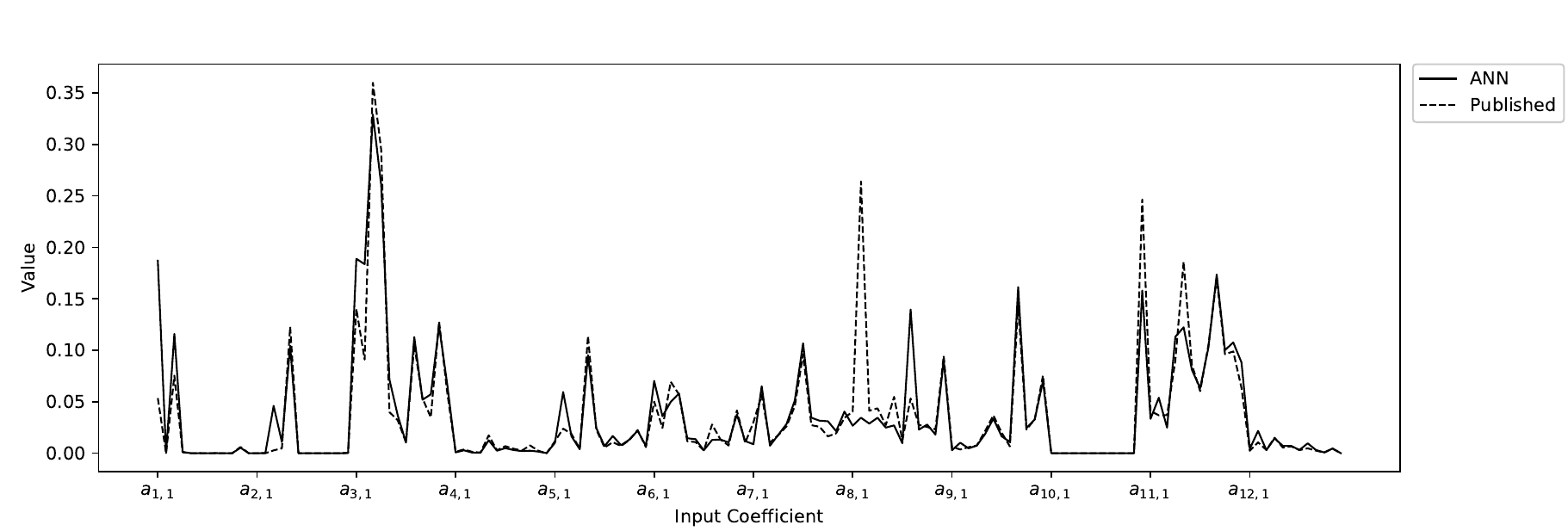}
        \subcaption{Input coefficients of Sapporo}\label{Fig:Forecasts01100}
    \end{minipage}\\
    \begin{minipage}[c]{0.95\hsize}
        \centering
        \includegraphics[width=0.9\columnwidth, clip]{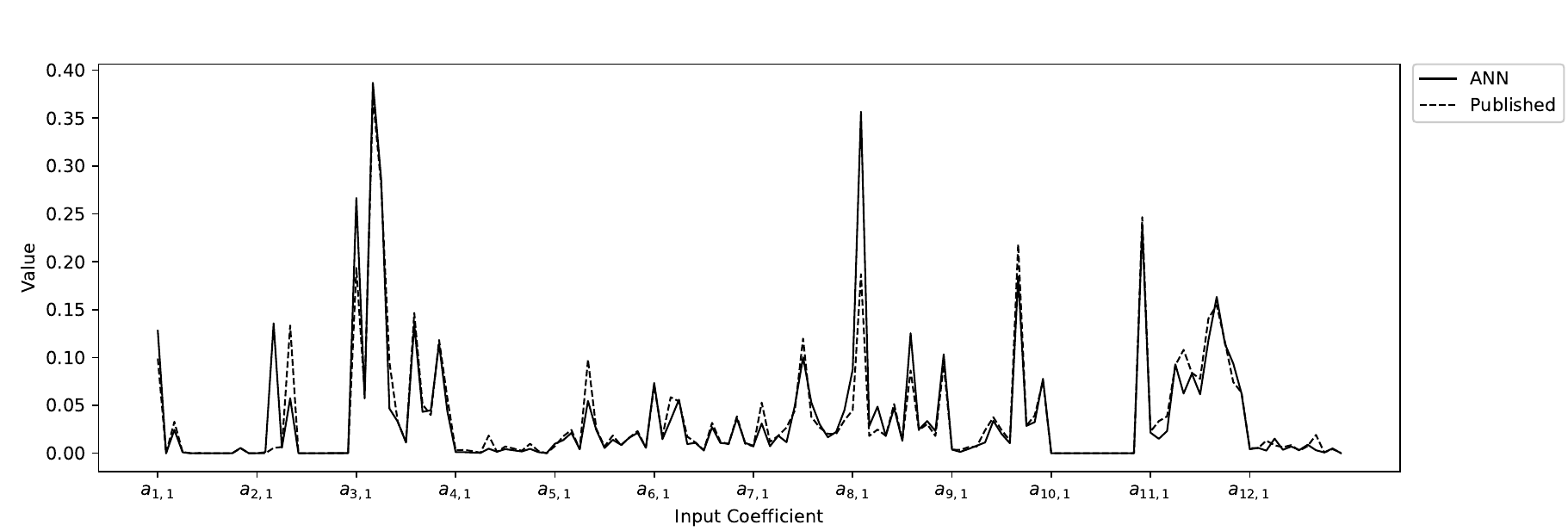}
        \subcaption{Input coefficients of Okayama}\label{Fig:Forecasts33100}
    \end{minipage}
    \caption{Predicted input coefficients for three cities using deep learning}
\end{figure}
Figures \ref{Fig:Forecasts21219}, \ref{Fig:Forecasts01100}, and \ref{Fig:Forecasts33100} show the deep learning predictions (ANN) and published city estimates (Published) for the cities of Gujo, Sapporo, and Okayama, respectively.
In all figures, the input coefficients are listed on the horizontal axis in the order $a_{1,1}, a_{1,2}, \ldots, a_{12,11}, a_{12,12}$.
These figures are prepared with reference to \cite{Papadas2002}.

For the input coefficients of the three cities, the ANN estimates are generally close to the city's published values, but for some input coefficients, such as $a_{1,1}$ and $a_{8,2}$, the estimates and published values differ substantially.
The difference is supposed to be due to the fact that the ANN does not fully capture the additional information used by the city in the process of calculating the published values.
At the same time, it should be noted that the published values of the cities also contain estimation errors relative to the true values, since they are in fact estimates.

\section{Discussion}
In terms of the results of predicting input coefficients using the method of this study for Japan as a whole, the prediction by deep learning is capable of high and stable accuracy compared to the conventional methods.
In addition, the assumptions required for the method in this study are less than those required for LQ and RAS.
LQ requires the selection of variables to be used for the location quotient, and FLQ further needs the setting or estimation of parameters.
RAS also requires the establishment of an initial input coefficient matrix and the estimation of total intermediate demands, total intermediate inputs, and total gross outputs for the region of interest.
On the other hand, the method of this research requires only the prior knowledge necessary for mixup.

The trained model in this study can be viewed as an integration of the various methods used to estimate the input--output table for each prefecture and city.
Most of the input--output tables published by the prefectures are estimated using available primary data and independent surveys according to general guidelines set by each prefecture.
In the case of the cities, input--output tables are estimated using different methods for each city, such as hybrid and non-survey methods, because the primary data for constructing input--output tables are extremely limited.
\footnote{
    However, the detailed estimation methods of input-output tables are rarely explained in these areas of Japan.
}
Since this study assumes a single model $f_{i,j} (\boldsymbol{x})$ for the input coefficient estimation method, the trained model is a synthesis of the estimation methods for each region.
In this sense, the methodology of this study has a meta-analytic aspect.

In the field of machine learning, various models are developed and improved to improve the prediction accuracy.
In this study, a very simple ANN is used as the prediction model, but there is room to further improve the prediction accuracy of input coefficients by applying more advanced models.
It should be noted that deep learning requires a large amount of time for model training.
In this prediction, model learning is performed for each of the 131 input coefficients, and the computation time required for model learning and prediction is extremely long compared to conventional estimation methods.
However, the problem of computation time can be solved by hardware.
For example, if the models to be trained are distributed over $n$ computers, the computation time can in principle be reduced to $1/n$ compared to the case where only one computer is used.
Alternatively, the speed can be increased by using fast CPUs or GPUs.
Therefore, if sufficient computing resources are available, the computation time required for deep learning becomes a trivial problem.

The method of this study may be less accurate in forecasting for specific regions.
In large regions such as prefectures, resources are distributed among many industries, while in very small regions such as towns and villages, resources may be concentrated in specific industries.
Since mixup in this study generates a virtual region from prefectures and cities, the distribution of resources in this virtual region is similar to that of the original regions.
Thus, for cities, towns, and villages that are concentrated in a particular industry, the less vicinal the areas are to the region generated by mixup, the less accurate the learned model will be in predicting them.

Furthermore, the method of this research may produce forecasts with low accuracy for future points of time.
In fact, using this method, I trained an ANN with a 2011 data set and predicted input coefficients for 2015, and found that the accuracy was very low.
As with most econometric models, the situation at the time of measurement is reflected in the ANN through the data set.
Since economic conditions have changed between 2011 and 2015, it is difficult to predict input coefficients for 2015 with high accuracy using a trained model with a 2011 data set.

The approach in this study could be applied to the forecasting of economic data, such as price indexes, where the generation process can be expressed as a single function.
The following points should be considered in its application.

The original data set must be of a certain size to achieve a high degree of prediction accuracy.
Model learning with the data set generated by mixup is equivalent to preventing over-fitting by regularizing model learning with the original data set\citep{Wu2020}.
Therefore, if the original data set is too small, the prediction accuracy of the trained model will remain low even if the data set is augmented by mixup.

The model being trained is heavily influenced by outliers.
As with ordinary linear regression, deep learning is affected by outliers.
Moreover, due to the nature of mixup, one outlier is spread across many hypothetical observations, making the effect of outliers even more impactful.

In order to perform mixup, the source variables for explanatory and objective variables are limited to those that can be composited and scaled.
For the variables used in this analysis, such as population, number of firms, and income, composition can be performed as the sum of multiple regions and scaling can be performed as the product of a single ratio.
However, indicator variables such as interest rates cannot be subject to data augmentation by mixup because they are difficult to compose and scale directly.

In the application of mixup proposed in this study, it is necessary to establish the prior knowledge that ``for a virtual region obtained by linear interpolation, the feature vectors should lead to the associated target.''
If this prior knowledge is not satisfied, mixup cannot be performed directly.
For example, it is quite difficult to learn a production function that is not constant returns to scale by mixup.
Let $y_i$ be the output of region $i$ and $boldsymbol{x}_i$ be the production factor vector. If the production function $g$ is not constant returns to scale, then the prior knowledge required for mixup is not satisfied as follows:
\[
    \lambda_1 y_1 + \lambda_2 y_2 \neq g(\lambda_1 \boldsymbol{x}_1 + \lambda_2 \boldsymbol{x}_2).
\]
Another example is the wrong composition of regions. As explained earlier, mixup between a prefecture and the city it contains is incorrect.
For these regions, it is not possible to get a sum for values that include transfers, so the composition cannot be meaningful.
The same is true for regions between different points in time, and if we are going to do the composition, we need to consider the treatment of transfers between regions.

\section*{Acknowledgments}
The author is very grateful to Dr. Shinya Kato (Yamaguchi University) for suggestions on confirming the prediction accuracy of the method of this paper.

This preprint has not undergone peer review (when applicable) or any post-submission improvements or corrections. The Version of Record of this article is published in Computational Economics, and is available online at \url{https://doi.org/10.1007/s10614-024-10641-1}.

\section*{Statements and Declarations}
\subsection*{Funding} The author declares that no funds, grants, or other support were received during the preparation of this manuscript.
\subsection*{Competing Interests} The author has no relevant financial or non-financial interests to disclose.

\begin{appendices}
\section{On the sum of intermediate inputs and gross outputs between regions} \label{Sec:Appendix}
For intermediate inputs and gross outputs, we confirm that the sum of the values for the two regions is equal to the value when these regions are considered as one.
\footnote{
    The following explanation was originally presented in an article written by the author in 2021 in Japanese. I am translating it into English and restating it here.
}

Consider the creation of a single region $R12$, which is the composition of any two regions (denoted $R1$ and $R2$, respectively).
In the case of a regional input--output table of competitive import type, the intermediate input of $R12$ is equal to the sum of the intermediate inputs of $R1$ and $R2$.
The intermediate input in a regional input--output table of the competitive import type is the sum of the input between industries within the region, the receiving between industries from other regions of the country to the region, and the import between industries from abroad to the region\citep{Fujimoto2019}.
That is, for the intermediate input $m_{i,j}$ from industry $i$ to industry $j$ in a region, if $\hat{m}_{i,j}$ is the input within the region, $\dot{m}_{i,j}$ is the receiving from other regions in Japan to industry $j$ of the region, and $\tilde{m}_{i,j}$ is the import from abroad to industry $j$ of the region, then
\[
    m_{i,j} = \hat{m}_{i,j} + \dot{m}_{i,j} + \tilde{m}_{i,j}.
\]
These intermediate inputs in $R1$ are denoted by $(m^{R1}_{i,j}, \hat{m}^{R1}_{i,j}, \dot{m}^{R1}_{i,j}, \tilde{m}^{R1}_{i,j})$. Similarly, for $R2$, denote $(m^{R2}_{i,j}, \dot{m}^{R2}_{i,j}, \dot{m}^{R2}_{i,j}, \tilde{m}^{R2}_{i,j})$.
Adding the intermediate inputs for these two regions yields
\[
    m^{R1}_{i,j} + m^{R2}_{i,j} = \hat{m}^{R1}_{i,j} + \dot{m}^{R1}_{i,j} + \tilde{m}^{R1}_{i,j} + \hat{m}^{R2}_{i,j} + \dot{m}^{R2}_{i,j} + \tilde{m}^{R2}_{i,j}.
\]
Inputs within $R12$($\hat{m}^{R12}_{i,j}$), inputs from other regions in the country to $R12$($\dot{m}^{R12}_{i,j}$), and foreign inputs to $R12$($\tilde{m}^{R12}_{i,j}$) are
\begin{align*}
    \hat{m}^{R12}_{i,j} &= \hat{m}^{R1}_{i,j} + \hat{m}^{R2}_{i,j} + (\text{input from} \ R2 \ \text{out of} \ \dot{m}^{R1}_{i,j}) \\
    &\quad + (\text{input from} \ R1 \ \text{out of} \ \dot{m}^{R2}_{i,j}) \\
    \dot{m}^{R12}_{i,j} &= (\text{input from all other regions of the country except} \ R2 \ \text{out of} \ \dot{m}^{R1}_{i,j}) \\
    &\quad + (\text{input from all other regions of the country except} \ R1 \ \text{out of} \ \dot{m}^{R2}_{i,j}) \\
    \tilde{m}^{R12}_{i,j} &= \tilde{m}^{R1}_{i,j} + \tilde{m}^{R2}_{i,j}.
\end{align*}
The intermediate input $m^{R12}_{i,j}$ of $R12$ is the sum of $\hat{m}^{R12}_{i,j}$, $\dot{m}^{R12}_{i,j}$ and $\tilde{m}^{R12}_{i,j}$, so
\begin{align*}
    m^{R12}_{i,j} &= \hat{m}^{R12}_{i,j} + \dot{m}^{R12}_{i,j} + \tilde{m}^{R12}_{i,j} \\
     &= \hat{m}^{R1}_{i,j} + \hat{m}^{R2}_{i,j} \\
     &\quad + (\text{input from} \ R2 \ \text{out of} \ \dot{m}^{R1}_{i,j}) \\
     &\quad + (\text{input from} \ R1 \ \text{out of} \ \dot{m}^{R2}_{i,j}) \\
     &\quad + (\text{input from all other regions of the country except} \ R2 \ \text{out of} \ \dot{m}^{R1}_{i,j}) \\
     &\quad + (\text{input from all other regions of the country except} \ R1 \ \text{out of} \ \dot{m}^{R2}_{i,j}) \\
     &\quad + \tilde{m}^{R1}_{i,j} + \tilde{m}^{R2}_{i,j} \\
     &= \hat{m}^{R1}_{i,j} + \hat{m}^{R2}_{i,j} + \dot{m}^{R1}_{i,j} + \dot{m}^{R2}_{i,j} + \tilde{m}^{R1}_{i,j} + \tilde{m}^{R2}_{i,j} \\
     &= m^{R1}_{i,j} + m^{R2}_{i,j}.
\end{align*}
Therefore, the sum of intermediate inputs for any two regions is equal to the intermediate input when these two regions are aggregated and considered as one new region.

We confirm that the gross output $Y^{R12}_i$ of industry $i$ in $R12$ is equal to the sum of the gross output of $R1$ and $R2$, $Y^{R1}_i + Y^{R2}_i$.
In the input--output table of competitive import type, as well as intermediate inputs, final demand for $R12$ can also be calculated as the sum of final demand for $R1$ and $R2$.
Since exports in $R12$ are equal to the sum of exports in $R1$ and $R2$, and imports in $R12$ are similar, net exports (the difference between exports and imports) in $R12$ are equal to the sum of net exports in $R1$ and $R2$.
The shipping to the other regions of industry $i$ in $R12$, $L^{R12}_i$, is as follows:
\begin{align*}
    L^{R12}_i &= L^{R1}_i + L^{R2}_i - \sum_j (\text{input from} \ R1 \ \text{out of} \ \dot{m}^{R2}_{i,j}) \\
    &\quad - (\text{input from} \ R1 \ \text{out of} \ F^{R2}_i) \\
    &\quad - \sum_j (\text{input from} \ R2 \ \text{out of} \ \dot{m}^{R1}_{i,j}) \\
    &\quad - (\text{input from} \ R2 \ \text{out of} \ F^{R1}_i).
\end{align*}
where $F^{R1}_i$ is the final demand for industry $i$ in $R1$ and $F^{R2}_i$ is the final demand for industry $i$ in $R2$.
The receiving from the other regions of industry $i$ in $R12$, $N^{R12}_i$, is calculated in the same way as for $L^{R12}_i$. Thus,
\begin{align*}
    N^{R12}_i &= N^{R1}_i + N^{R2}_i - \sum_j (\text{input from} \ R2 \ \text{out of} \ \dot{m}^{R1}_{i,j}) \\
    &\quad - (\text{input from} \ R2 \ \text{out of} \ F^{R1}_i) \\
    &\quad - \sum_j (\text{input from} \ R1 \ \text{out of} \ \dot{m}^{R2}_{i,j}) \\
    &\quad - (\text{input from} \ R1 \ \text{out of} \ F^{R2}_i).
\end{align*}
From these equations, the net transfer of industry $i$ in $R12$ is equal to the sum of the net transfers in $R1$ and $R2$ as follows:
\[
    L^{R12}_i - N^{R12}_i = L^{R1}_i + L^{R2}_i - N^{R1}_i - N^{R2}_i.
\]
Gross output is the sum of total intermediate demand ($=$ the row sum of intermediate inputs), final demand, net transfers, and net exports.
Thus, output $Y^{R12}_i$ in $R12$ can be calculated as $Y^{R1}_i + Y^{R2}_i$.

\end{appendices}


\begin{thebibliography}{}
    \renewcommand{\doi}[1]{\url{https://doi.org/#1}}
    \bibcommenthead

    \bibitem [\protect \citeauthoryear {%
    Abbasimehr%
    , Shabani%
    \BCBL {}\ \BBA {} Mohsen%
    }{%
    Abbasimehr%
    \ \protect \BOthers {.}}{%
    {\protect \APACyear {2020}}%
    }]{%
    Abbasimehr2020}
    \APACinsertmetastar {%
    Abbasimehr2020}%
    \begin{APACrefauthors}%
    Abbasimehr, H.%
    , Shabani, M.%
    \BCBL {} Mohsen, Y.%
    \end{APACrefauthors}%
    \unskip\
    \newblock
    \APACrefYearMonthDay{2020}{}{}.
    \newblock
    {\BBOQ}\APACrefatitle {An Optimized Model Using {LSTM} Network for Demand
      Forecasting} {An optimized model using {LSTM} network for demand
      forecasting}.{\BBCQ}
    \newblock
    \APACjournalVolNumPages{Computers \& Industrial Engineering}{143}{}{106435,}
    \newblock
    \begin{APACrefDOI} \doi{10.1016/j.cie.2020.106435} \end{APACrefDOI}
    \newblock

    \newblock

    \PrintBackRefs{\CurrentBib}

    \bibitem [\protect \citeauthoryear {%
    Bacharach%
    }{%
    Bacharach%
    }{%
    {\protect \APACyear {1970}}%
    }]{%
    Bacharach1970}
    \APACinsertmetastar {%
    Bacharach1970}%
    \begin{APACrefauthors}%
    Bacharach, M.%
    \end{APACrefauthors}%
    \unskip\
    \newblock
    \APACrefYear{1970}.
    \newblock
    \APACrefbtitle {Biproportional Matrices \& Input--Output Change}
      {Biproportional matrices \& input--output change}.
    \newblock
    \APACaddressPublisher{}{Cambridge University Press}.
    \PrintBackRefs{\CurrentBib}

    \bibitem [\protect \citeauthoryear {%
    Bonfiglio%
    \ \BBA {} Francesco%
    }{%
    Bonfiglio%
    \ \BBA {} Francesco%
    }{%
    {\protect \APACyear {2008}}%
    }]{%
    Bonfiglio2008}
    \APACinsertmetastar {%
    Bonfiglio2008}%
    \begin{APACrefauthors}%
    Bonfiglio, A.%
    \BCBT {}\ \BBA {} Francesco, C.%
    \end{APACrefauthors}%
    \unskip\
    \newblock
    \APACrefYearMonthDay{2008}{}{}.
    \newblock
    {\BBOQ}\APACrefatitle {Assessing the Behaviour of Non-Survey Methods for
      Constructing Regional Input--Output Tables through a {M}onte {C}arlo
      Simulation} {Assessing the behaviour of non-survey methods for constructing
      regional input--output tables through a {M}onte {C}arlo simulation}.{\BBCQ}
    \newblock
    \APACjournalVolNumPages{Economic Systems Research}{20}{3}{243--258,}
    \newblock
    \begin{APACrefDOI} \doi{10.1080/09535310802344315} \end{APACrefDOI}
    \newblock

    \newblock

    \PrintBackRefs{\CurrentBib}

    \bibitem [\protect \citeauthoryear {%
    Botev%
    , Lever%
    \BCBL {}\ \BBA {} Barber%
    }{%
    Botev%
    \ \protect \BOthers {.}}{%
    {\protect \APACyear {2017}}%
    }]{%
    Botev2017}
    \APACinsertmetastar {%
    Botev2017}%
    \begin{APACrefauthors}%
    Botev, A.%
    , Lever, G.%
    \BCBL {} Barber, D.%
    \end{APACrefauthors}%
    \unskip\
    \newblock
    \APACrefYearMonthDay{2017}{}{}.
    \newblock
    {\BBOQ}\APACrefatitle {{N}esterov's Accelerated Gradient and Momentum as
      Approximations to Regularised Update Descent} {{N}esterov's accelerated
      gradient and momentum as approximations to regularised update
      descent}.{\BBCQ}
    \newblock
     \APACrefbtitle {2017 International Joint Conference on Neural Networks} {2017
      international joint conference on neural networks}\ (\BPGS\ 1899--1903).
    \PrintBackRefs{\CurrentBib}

    \bibitem [\protect \citeauthoryear {%
    Chapelle%
    , Weston%
    , Bottou%
    \BCBL {}\ \BBA {} Vapnik%
    }{%
    Chapelle%
    \ \protect \BOthers {.}}{%
    {\protect \APACyear {2000}}%
    }]{%
    Chapelle2000}
    \APACinsertmetastar {%
    Chapelle2000}%
    \begin{APACrefauthors}%
    Chapelle, O.%
    , Weston, J.%
    , Bottou, L.%
    \BCBL {} Vapnik, V.%
    \end{APACrefauthors}%
    \unskip\
    \newblock
    \APACrefYearMonthDay{2000}{}{}.
    \newblock
    {\BBOQ}\APACrefatitle {Vicinal Risk Minimization} {Vicinal risk
      minimization}.{\BBCQ}
    \newblock
     T.~Leen, T.~Dietterich\BCBL {}\ \BBA {} V.~Tresp\ (\BEDS), \APACrefbtitle
      {Advances in Neural Information Processing Systems} {Advances in neural
      information processing systems}\ (\BVOL~13, \BPGS\ 416--422).
    \newblock
    \APACaddressPublisher{}{MIT Press}.
    \PrintBackRefs{\CurrentBib}

    \bibitem [\protect \citeauthoryear {%
    Dao%
    \ \protect \BOthers {.}}{%
    Dao%
    \ \protect \BOthers {.}}{%
    {\protect \APACyear {2019}}%
    }]{%
    Dao2019}
    \APACinsertmetastar {%
    Dao2019}%
    \begin{APACrefauthors}%
    Dao, T.%
    , Gu, A.%
    , Ratner, A.J.%
    , Virginia, S.%
    , De~Sa, C.%
    \BCBL {} R\'{e}, C.%
    \end{APACrefauthors}%
    \unskip\
    \newblock
    \APACrefYearMonthDay{2019}{}{}.
    \newblock
    {\BBOQ}\APACrefatitle {A Kernel Theory of Modern Data Augmentation} {A kernel
      theory of modern data augmentation}.{\BBCQ}
    \newblock
     \APACrefbtitle {International Conference on Machine Learning} {International
      conference on machine learning}\ (\BPGS\ 1528--1537).
    \PrintBackRefs{\CurrentBib}

    \bibitem [\protect \citeauthoryear {%
    Flegg%
    , Lamonica%
    , Chelli%
    , Recchioni%
    \BCBL {}\ \BBA {} Tohmo%
    }{%
    Flegg%
    \ \protect \BOthers {.}}{%
    {\protect \APACyear {2021}}%
    }]{%
    Flegg2021}
    \APACinsertmetastar {%
    Flegg2021}%
    \begin{APACrefauthors}%
    Flegg, A.T.%
    , Lamonica, G.R.%
    , Chelli, F.M.%
    , Recchioni, M.C.%
    \BCBL {} Tohmo, T.%
    \end{APACrefauthors}%
    \unskip\
    \newblock
    \APACrefYearMonthDay{2021}{}{}.
    \newblock
    {\BBOQ}\APACrefatitle {A New Approach to Modelling the Input--Output Structure
      of Regional Economies Using Non-Survey Methods} {A new approach to modelling
      the input--output structure of regional economies using non-survey
      methods}.{\BBCQ}
    \newblock
    \APACjournalVolNumPages{Journal of Economic Structures}{10}{12}{,}
    \newblock
    \begin{APACrefDOI} \doi{10.1186/s40008-021-00242-8} \end{APACrefDOI}
    \newblock

    \newblock

    \PrintBackRefs{\CurrentBib}

    \bibitem [\protect \citeauthoryear {%
    Flegg%
    \ \BBA {} Tohmo%
    }{%
    Flegg%
    \ \BBA {} Tohmo%
    }{%
    {\protect \APACyear {2013}}%
    }]{%
    Flegg2013}
    \APACinsertmetastar {%
    Flegg2013}%
    \begin{APACrefauthors}%
    Flegg, A.T.%
    \BCBT {}\ \BBA {} Tohmo, T.%
    \end{APACrefauthors}%
    \unskip\
    \newblock
    \APACrefYearMonthDay{2013}{}{}.
    \newblock
    \APACrefbtitle {Estimating Regional Input Coefficients and Multipliers: The Use
      of the {FLQ} Is Not a Gamble} {Estimating regional input coefficients and
      multipliers: The use of the {FLQ} is not a gamble}\ \APACbVolEdTR{}{\BTR{}}.
    \newblock
    \APACaddressInstitution{}{University of the West of England}.
    \PrintBackRefs{\CurrentBib}

    \bibitem [\protect \citeauthoryear {%
    Fujimoto%
    }{%
    Fujimoto%
    }{%
    {\protect \APACyear {2019}}%
    }]{%
    Fujimoto2019}
    \APACinsertmetastar {%
    Fujimoto2019}%
    \begin{APACrefauthors}%
    Fujimoto, T.%
    \end{APACrefauthors}%
    \unskip\
    \newblock
    \APACrefYearMonthDay{2019}{}{}.
    \newblock
    {\BBOQ}\APACrefatitle {Appropriate Assumption on Cross-hauling National
      Input--Output Table Regionalization} {Appropriate assumption on cross-hauling
      national input--output table regionalization}.{\BBCQ}
    \newblock
    \APACjournalVolNumPages{Spatial Economic Analysis}{14}{1}{106--128,}
    \newblock
    \begin{APACrefDOI} \doi{10.1080/17421772.2018.1506151} \end{APACrefDOI}
    \newblock

    \newblock

    \PrintBackRefs{\CurrentBib}

    \bibitem [\protect \citeauthoryear {%
    Gerking%
    }{%
    Gerking%
    }{%
    {\protect \APACyear {1976}}%
    }]{%
    Gerking1976}
    \APACinsertmetastar {%
    Gerking1976}%
    \begin{APACrefauthors}%
    Gerking, S.D.%
    \end{APACrefauthors}%
    \unskip\
    \newblock
    \APACrefYear{1976}.
    \newblock
    \APACrefbtitle {Estimation of Stochastic Input--Output Models: Some Statistical
      Problems} {Estimation of stochastic input--output models: Some statistical
      problems}.
    \newblock
    \APACaddressPublisher{}{Springer}.
    \PrintBackRefs{\CurrentBib}

    \bibitem [\protect \citeauthoryear {%
    G\'{e}ron%
    }{%
    G\'{e}ron%
    }{%
    {\protect \APACyear {2019}}%
    }]{%
    Geron2019}
    \APACinsertmetastar {%
    Geron2019}%
    \begin{APACrefauthors}%
    G\'{e}ron, A.%
    \end{APACrefauthors}%
    \unskip\
    \newblock
    \APACrefYear{2019}.
    \newblock
    \APACrefbtitle {Hands-On Machine Learning with Scikit-Learn, {K}eras, and
      {T}ensor{F}low: Concepts, Tools, and Techniques to Build Intelligent Systems}
      {Hands-on machine learning with scikit-learn, {K}eras, and {T}ensor{F}low:
      Concepts, tools, and techniques to build intelligent systems}.
    \newblock
    \APACaddressPublisher{}{O'Reilly}.
    \PrintBackRefs{\CurrentBib}

    \bibitem [\protect \citeauthoryear {%
    He%
    , Zhang%
    , Ren%
    \BCBL {}\ \BBA {} Sun%
    }{%
    He%
    \ \protect \BOthers {.}}{%
    {\protect \APACyear {2015}}%
    }]{%
    He2015}
    \APACinsertmetastar {%
    He2015}%
    \begin{APACrefauthors}%
    He, K.%
    , Zhang, X.%
    , Ren, S.%
    \BCBL {} Sun, J.%
    \end{APACrefauthors}%
    \unskip\
    \newblock
    \APACrefYearMonthDay{2015}{}{}.
    \newblock
    {\BBOQ}\APACrefatitle {Delving Deep into Rectifiers: Surpassing Human--Level
      Performance on {I}mage{N}et Classification} {Delving deep into rectifiers:
      Surpassing human--level performance on {I}mage{N}et classification}.{\BBCQ}
    \newblock
     \APACrefbtitle {Proceedings of the {IEEE} International Conference on Computer
      Vision} {Proceedings of the {IEEE} international conference on computer
      vision}\ (\BPGS\ 1026--1034).
    \PrintBackRefs{\CurrentBib}

    \bibitem [\protect \citeauthoryear {%
    Hewings%
    }{%
    Hewings%
    }{%
    {\protect \APACyear {1977}}%
    }]{%
    Hewings1977}
    \APACinsertmetastar {%
    Hewings1977}%
    \begin{APACrefauthors}%
    Hewings, G.J.D.%
    \end{APACrefauthors}%
    \unskip\
    \newblock
    \APACrefYearMonthDay{1977}{}{}.
    \newblock
    {\BBOQ}\APACrefatitle {Evaluating the Possibilities for Exchanging Regional
      Input--Output Coefficients} {Evaluating the possibilities for exchanging
      regional input--output coefficients}.{\BBCQ}
    \newblock
    \APACjournalVolNumPages{Environment and Planning A}{9}{8}{927--944,}
    \newblock
    \begin{APACrefDOI} \doi{10.1068/a090927} \end{APACrefDOI}
    \newblock

    \newblock

    \PrintBackRefs{\CurrentBib}

    \bibitem [\protect \citeauthoryear {%
    Hol{\`y}%
    \ \BBA {} {\v{S}}afr%
    }{%
    Hol{\`y}%
    \ \BBA {} {\v{S}}afr%
    }{%
    {\protect \APACyear {2022}}%
    }]{%
    Holy2022}
    \APACinsertmetastar {%
    Holy2022}%
    \begin{APACrefauthors}%
    Hol{\`y}, V.%
    \BCBT {}\ \BBA {} {\v{S}}afr, K.%
    \end{APACrefauthors}%
    \unskip\
    \newblock
    \APACrefYearMonthDay{2022}{}{}.
    \newblock
    {\BBOQ}\APACrefatitle {Disaggregating Input--Output Tables by the
      Multidimensional {RAS} Method: A Case Study of the {C}zech {R}epublic}
      {Disaggregating input--output tables by the multidimensional {RAS} method: A
      case study of the {C}zech {R}epublic}.{\BBCQ}
    \newblock
    \APACjournalVolNumPages{Economic Systems Research}{}{}{1--23,}
    \newblock
    \begin{APACrefDOI} \doi{10.1080/09535314.2022.2091978} \end{APACrefDOI}
    \newblock

    \newblock

    \PrintBackRefs{\CurrentBib}

    \bibitem [\protect \citeauthoryear {%
    Hosoe%
    }{%
    Hosoe%
    }{%
    {\protect \APACyear {2014}}%
    }]{%
    Hosoe2013}
    \APACinsertmetastar {%
    Hosoe2013}%
    \begin{APACrefauthors}%
    Hosoe, N.%
    \end{APACrefauthors}%
    \unskip\
    \newblock
    \APACrefYearMonthDay{2014}{}{}.
    \newblock
    {\BBOQ}\APACrefatitle {Estimation Errors in Input-Output Tables and Prediction
      Errors in Computable General Equilibrium Analysis} {Estimation errors in
      input-output tables and prediction errors in computable general equilibrium
      analysis}.{\BBCQ}
    \newblock
    \APACjournalVolNumPages{Economic Modelling}{}{}{277--286,}
    \newblock
    \begin{APACrefDOI} \doi{10.1016/j.econmod.2014.07.012} \end{APACrefDOI}
    \newblock

    \newblock

    \PrintBackRefs{\CurrentBib}

    \bibitem [\protect \citeauthoryear {%
    Isserman%
    }{%
    Isserman%
    }{%
    {\protect \APACyear {1977}}%
    }]{%
    Isserman1977}
    \APACinsertmetastar {%
    Isserman1977}%
    \begin{APACrefauthors}%
    Isserman, A.M.%
    \end{APACrefauthors}%
    \unskip\
    \newblock
    \APACrefYearMonthDay{1977}{}{}.
    \newblock
    {\BBOQ}\APACrefatitle {The Location Quotient Approach to Estimating Regional
      Economic Impacts} {The location quotient approach to estimating regional
      economic impacts}.{\BBCQ}
    \newblock
    \APACjournalVolNumPages{Journal of the American Institute of
      Planners}{43}{1}{33--41,}
    \newblock
    \begin{APACrefDOI} \doi{10.1080/01944367708977758} \end{APACrefDOI}
    \newblock

    \newblock

    \PrintBackRefs{\CurrentBib}

    \bibitem [\protect \citeauthoryear {%
    Lahr%
    \ \BBA {} De~Mesnard%
    }{%
    Lahr%
    \ \BBA {} De~Mesnard%
    }{%
    {\protect \APACyear {2004}}%
    }]{%
    Lahr2004}
    \APACinsertmetastar {%
    Lahr2004}%
    \begin{APACrefauthors}%
    Lahr, M.L.%
    \BCBT {}\ \BBA {} De~Mesnard, L.%
    \end{APACrefauthors}%
    \unskip\
    \newblock
    \APACrefYearMonthDay{2004}{}{}.
    \newblock
    {\BBOQ}\APACrefatitle {Biproportional Techniques in Input--Output Analysis:
      Table Updating and Structural Analysis} {Biproportional techniques in
      input--output analysis: Table updating and structural analysis}.{\BBCQ}
    \newblock
    \APACjournalVolNumPages{Economic Systems Research}{16}{2}{115--134,}
    \newblock
    \begin{APACrefDOI} \doi{10.1080/0953531042000219259} \end{APACrefDOI}
    \newblock

    \newblock

    \PrintBackRefs{\CurrentBib}

    \bibitem [\protect \citeauthoryear {%
    Law%
    , Li%
    , Fong%
    \BCBL {}\ \BBA {} Han%
    }{%
    Law%
    \ \protect \BOthers {.}}{%
    {\protect \APACyear {2019}}%
    }]{%
    Law2019}
    \APACinsertmetastar {%
    Law2019}%
    \begin{APACrefauthors}%
    Law, R.%
    , Li, G.%
    , Fong, D.K.C.%
    \BCBL {} Han, X.%
    \end{APACrefauthors}%
    \unskip\
    \newblock
    \APACrefYearMonthDay{2019}{}{}.
    \newblock
    {\BBOQ}\APACrefatitle {Tourism Demand Forecasting: A Deep Learning Approach}
      {Tourism demand forecasting: A deep learning approach}.{\BBCQ}
    \newblock
    \APACjournalVolNumPages{Annals of Tourism Research}{75}{}{410--423,}
    \newblock
    \begin{APACrefDOI} \doi{10.1016/j.annals.2019.01.014} \end{APACrefDOI}
    \newblock

    \newblock

    \PrintBackRefs{\CurrentBib}

    \bibitem [\protect \citeauthoryear {%
    Papadas%
    \ \BBA {} Hutchinson%
    }{%
    Papadas%
    \ \BBA {} Hutchinson%
    }{%
    {\protect \APACyear {2002}}%
    }]{%
    Papadas2002}
    \APACinsertmetastar {%
    Papadas2002}%
    \begin{APACrefauthors}%
    Papadas, C.T.%
    \BCBT {}\ \BBA {} Hutchinson, W.G.%
    \end{APACrefauthors}%
    \unskip\
    \newblock
    \APACrefYearMonthDay{2002}{}{}.
    \newblock
    {\BBOQ}\APACrefatitle {Neural Network Forecasts of Input-Output Technology}
      {Neural network forecasts of input-output technology}.{\BBCQ}
    \newblock
    \APACjournalVolNumPages{Applied Economics}{34}{}{1607--1615,}
    \newblock
    \begin{APACrefDOI} \doi{10.1080/00036840110118133} \end{APACrefDOI}
    \newblock

    \newblock

    \PrintBackRefs{\CurrentBib}

    \bibitem [\protect \citeauthoryear {%
    Ramyar%
    \ \BBA {} Kianfar%
    }{%
    Ramyar%
    \ \BBA {} Kianfar%
    }{%
    {\protect \APACyear {2019}}%
    }]{%
    Ramyar2019}
    \APACinsertmetastar {%
    Ramyar2019}%
    \begin{APACrefauthors}%
    Ramyar, S.%
    \BCBT {}\ \BBA {} Kianfar, F.%
    \end{APACrefauthors}%
    \unskip\
    \newblock
    \APACrefYearMonthDay{2019}{}{}.
    \newblock
    {\BBOQ}\APACrefatitle {Forecasting Crude Oil Prices: A Comparison Between
      Artificial Neural Networks and Vector Autoregressive Models} {Forecasting
      crude oil prices: A comparison between artificial neural networks and vector
      autoregressive models}.{\BBCQ}
    \newblock
    \APACjournalVolNumPages{Computational Economics}{53}{2}{743--761,}
    \newblock
    \begin{APACrefDOI} \doi{10.1007/s10614-017-9764-7} \end{APACrefDOI}
    \newblock

    \newblock

    \PrintBackRefs{\CurrentBib}

    \bibitem [\protect \citeauthoryear {%
    Richardson%
    }{%
    Richardson%
    }{%
    {\protect \APACyear {1985}}%
    }]{%
    Richardson1985}
    \APACinsertmetastar {%
    Richardson1985}%
    \begin{APACrefauthors}%
    Richardson, H.W.%
    \end{APACrefauthors}%
    \unskip\
    \newblock
    \APACrefYearMonthDay{1985}{}{}.
    \newblock
    {\BBOQ}\APACrefatitle {Input--Output and Economic Base Multipliers: Looking
      Backward and Forward} {Input--output and economic base multipliers: Looking
      backward and forward}.{\BBCQ}
    \newblock
    \APACjournalVolNumPages{Journal of Regional Science}{25}{4}{607--661,}
    \newblock
    \begin{APACrefDOI} \doi{10.1111/j.1467-9787.1985.tb00325.x} \end{APACrefDOI}
    \newblock

    \newblock

    \PrintBackRefs{\CurrentBib}

    \bibitem [\protect \citeauthoryear {%
    Riddington%
    , Gibson%
    \BCBL {}\ \BBA {} Anderson%
    }{%
    Riddington%
    \ \protect \BOthers {.}}{%
    {\protect \APACyear {2006}}%
    }]{%
    Riddington2006}
    \APACinsertmetastar {%
    Riddington2006}%
    \begin{APACrefauthors}%
    Riddington, G.%
    , Gibson, H.%
    \BCBL {} Anderson, J.%
    \end{APACrefauthors}%
    \unskip\
    \newblock
    \APACrefYearMonthDay{2006}{}{}.
    \newblock
    {\BBOQ}\APACrefatitle {Comparison of Gravity Model, Survey and Location
      Quotient-based Local Area Tables and Multipliers} {Comparison of gravity
      model, survey and location quotient-based local area tables and
      multipliers}.{\BBCQ}
    \newblock
    \APACjournalVolNumPages{Regional Studies}{40}{9}{1069--1081,}
    \newblock
    \begin{APACrefDOI} \doi{10.1080/00343400601047374} \end{APACrefDOI}
    \newblock

    \newblock

    \PrintBackRefs{\CurrentBib}

    \bibitem [\protect \citeauthoryear {%
    Round%
    }{%
    Round%
    }{%
    {\protect \APACyear {1983}}%
    }]{%
    Round1983}
    \APACinsertmetastar {%
    Round1983}%
    \begin{APACrefauthors}%
    Round, J.I.%
    \end{APACrefauthors}%
    \unskip\
    \newblock
    \APACrefYearMonthDay{1983}{}{}.
    \newblock
    {\BBOQ}\APACrefatitle {Nonsurvey Techniques: A Critical Review of the Theory
      and the Evidence} {Nonsurvey techniques: A critical review of the theory and
      the evidence}.{\BBCQ}
    \newblock
    \APACjournalVolNumPages{International Regional Science Review}{8}{3}{189--212,}
    \newblock
    \begin{APACrefDOI} \doi{10.1177/016001768300800302} \end{APACrefDOI}
    \newblock

    \newblock

    \PrintBackRefs{\CurrentBib}

    \bibitem [\protect \citeauthoryear {%
    Smith%
    }{%
    Smith%
    }{%
    {\protect \APACyear {2017}}%
    }]{%
    Smith2017}
    \APACinsertmetastar {%
    Smith2017}%
    \begin{APACrefauthors}%
    Smith, L.N.%
    \end{APACrefauthors}%
    \unskip\
    \newblock
    \APACrefYearMonthDay{2017}{}{}.
    \newblock
    {\BBOQ}\APACrefatitle {Cyclical Learning Rates for Training Neural Networks}
      {Cyclical learning rates for training neural networks}.{\BBCQ}
    \newblock
    \APACjournalVolNumPages{arXiv}{}{}{},
    \newblock
    {\href{https://arxiv.org/abs/1506.01186}{{arXiv:1506.01186}}}
    \newblock
     {[cs.CV]}
    \PrintBackRefs{\CurrentBib}

    \bibitem [\protect \citeauthoryear {%
    Szab\'{o}%
    }{%
    Szab\'{o}%
    }{%
    {\protect \APACyear {2015}}%
    }]{%
    Szabo2015}
    \APACinsertmetastar {%
    Szabo2015}%
    \begin{APACrefauthors}%
    Szab\'{o}, N.%
    \end{APACrefauthors}%
    \unskip\
    \newblock
    \APACrefYearMonthDay{2015}{}{}.
    \newblock
    {\BBOQ}\APACrefatitle {Methods for Regionalizing Input-Output Tables} {Methods
      for regionalizing input-output tables}.{\BBCQ}
    \newblock
    \APACjournalVolNumPages{Regional Statistics}{5}{1}{44--65,}
    \newblock
    \begin{APACrefDOI} \doi{10.15196/RS05103} \end{APACrefDOI}
    \newblock

    \newblock

    \PrintBackRefs{\CurrentBib}

    \bibitem [\protect \citeauthoryear {%
    Wu%
    , Zhang%
    , Valiant%
    \BCBL {}\ \BBA {} R\'{e}%
    }{%
    Wu%
    \ \protect \BOthers {.}}{%
    {\protect \APACyear {2020}}%
    }]{%
    Wu2020}
    \APACinsertmetastar {%
    Wu2020}%
    \begin{APACrefauthors}%
    Wu, S.%
    , Zhang, H.R.%
    , Valiant, G.%
    \BCBL {} R\'{e}, C.%
    \end{APACrefauthors}%
    \unskip\
    \newblock
    \APACrefYearMonthDay{2020}{}{}.
    \newblock
    {\BBOQ}\APACrefatitle {On the Generalization Effects of Linear Transformations
      in Data Augmentation} {On the generalization effects of linear
      transformations in data augmentation}.{\BBCQ}
    \newblock
    \APACjournalVolNumPages{arXiv}{}{}{},
    \newblock
    {\href{https://arxiv.org/abs/2005.00695}{{arXiv:2005.00695}}}
    \newblock
     {[cs.LG]}
    \PrintBackRefs{\CurrentBib}

    \bibitem [\protect \citeauthoryear {%
    H.~Zhang%
    , Cisse%
    , Dauphin%
    \BCBL {}\ \BBA {} Lopez-Paz%
    }{%
    H.~Zhang%
    \ \protect \BOthers {.}}{%
    {\protect \APACyear {2018}}%
    }]{%
    Zhang2018}
    \APACinsertmetastar {%
    Zhang2018}%
    \begin{APACrefauthors}%
    Zhang, H.%
    , Cisse, M.%
    , Dauphin, Y.N.%
    \BCBL {} Lopez-Paz, D.%
    \end{APACrefauthors}%
    \unskip\
    \newblock
    \APACrefYearMonthDay{2018}{}{}.
    \newblock
    {\BBOQ}\APACrefatitle {mixup: Beyond Empirical Risk Minimization} {mixup:
      Beyond empirical risk minimization}.{\BBCQ}
    \newblock
    \APACjournalVolNumPages{arXiv}{}{}{},
    \newblock
    {\href{https://arxiv.org/abs/1710.09412}{{arXiv:1710.09412}}}
    \newblock
     {[cs.LG]}
    \PrintBackRefs{\CurrentBib}

    \bibitem [\protect \citeauthoryear {%
    L.~Zhang%
    , Deng%
    , Kawaguchi%
    , Ghorbani%
    \BCBL {}\ \BBA {} Zou%
    }{%
    L.~Zhang%
    \ \protect \BOthers {.}}{%
    {\protect \APACyear {2021}}%
    }]{%
    Zhang2021}
    \APACinsertmetastar {%
    Zhang2021}%
    \begin{APACrefauthors}%
    Zhang, L.%
    , Deng, Z.%
    , Kawaguchi, K.%
    , Ghorbani, A.%
    \BCBL {} Zou, J.%
    \end{APACrefauthors}%
    \unskip\
    \newblock
    \APACrefYearMonthDay{2021}{}{}.
    \newblock
    {\BBOQ}\APACrefatitle {How Does Mixup Help with Robustness and Generalization?}
      {How does mixup help with robustness and generalization?}{\BBCQ}
    \newblock
    \APACjournalVolNumPages{arXiv}{}{}{},
    \newblock
    {\href{https://arxiv.org/abs/2010.04819}{{arXiv:2010.04819}}}
    \newblock
     {[cs.LG]}
    \PrintBackRefs{\CurrentBib}

\end{thebibliography}
\end{document}